\documentclass[fleqn,usenatbib]{mnras}

\usepackage{newtxtext,newtxmath}

\usepackage[T1]{fontenc}
\usepackage{ae,aecompl}

\usepackage{natbib}
\usepackage{journal_names}

\usepackage{graphicx}	
\usepackage{amsmath}	
\usepackage{amssymb}	
\usepackage{subfigure}  

\usepackage{booktabs}
\usepackage{longtable}

\usepackage{dirtytalk}
\usepackage[nolist,nohyperlinks]{acronym}
\usepackage[normalem]{ulem}

\usepackage{soul}

\title[Optical and radio study of the polar UZ Fornacis]{A spectroscopic, photometric, polarimetric and radio study of the eclipsing polar UZ Fornacis: the first simultaneous SALT and MeerKAT observations.}

\author[Z.N. Khangale et al.]{Z. N. Khangale,$^{1,2}$\thanks{E-mail: khangalezn@saao.ac.za (ZNK)} S. B. Potter,$^{1}$ P. A. Woudt,$^2$ D. A. H. Buckley,$^{1,3}$ A. N. Semena,$^{4}$\newauthor E. J. Kotze,$^{1,3}$ D. N. Groenewald,$^{1,3}$ D. M. Hewitt,$^{1,2}$ M. L. Pretorius,$^{1}$ R. P. 
Fender,$^{2,5}$ \newauthor P. Groot,$^{1,2,6}$ S. Bloemen,$^{6}$ M. Klein-Wolt,$^{6}$ E. K{\"o}rding,$^{6}$ R. Le Poole,$^{7}$ V. A. McBride,$^{1,2,8}$ \newauthor L. Townsend,$^{2}$ K. Paterson,$^{9}$ D. L. A. Pieterse$^{6}$ and P. Vreeswijk$^{6}$
\\
$^{1}$South African Astronomical Observatory, Observatory Road, Observatory, 7925, Cape Town, RSA\\
$^{2}$Inter-University Institute for Data-Intensive Astronomy, Department of Astronomy, University of Cape Town, Private Bag X3, \\ Rondebosch 7701, South Africa\\
$^{3}$Southern African Large Telescope, Observatory Road, Observatory, 7925, Cape Town, RSA\\
$^{4}$Space Research Institute, Russian Academy of Sciences, Profsoyuznaya 84/32, 117997, Moscow, Russia\\
$^{5}$Astrophysics, Department of Physics, University of Oxford, Keble Road, Oxford OX1 3RH, UK\\
$^{6}$Department of Astrophysics/IMAPP, Radboud University Nijmegen, PO Box 9010, NL-6500 GL Nijmegen, the Netherlands\\
$^{7}$Leiden Observatory, Leiden University, P.O. Box 9513, 2300 RA Leiden, the Netherlands\\
$^{8}$IAU-Office of Astronomy for Development, P.O. Box 9, 7935 Observatory, South Africa\\
$^{9}$ Center for Interdisciplinary Exploration and Research in Astrophysics (CIERA) and Department of Physics and Astronomy, \\ Northwestern University, 1800 Sherman Ave, Evanston, IL 60201, USA
}

\date{Accepted 09 January 2020. Received 13 December 2019 ; in original form 13 December 2019}

\pubyear{2020}

\begin{document}
\label{firstpage}
\pagerange{\pageref{firstpage}--\pageref{lastpage}}
\maketitle

\begin{abstract}
We present phase-resolved spectroscopy, photometry and circular spectropolarimetry of the eclipsing polar UZ Fornacis. Doppler tomography of the strongest emission lines using the inside-out projection revealed the presence of three emission regions: from the irradiated face of the secondary star, the ballistic stream and the threading region, and the magnetically confined accretion stream. 
The total intensity spectrum shows broad emission features and a continuum that rises in the blue. The circularly polarized spectrum shows the presence of three cyclotron emission harmonics at $\sim$4500 \AA{}, 6000 \AA{} and 7700 \AA{}, corresponding to harmonic numbers 4, 3, and 2, respectively. 
These features are dominant before the eclipse and disappear after the eclipse. The harmonics are consistent with a magnetic field strength of $\sim$57 MG. We also present phase-resolved circular and linear photopolarimetry to complement the spectropolarimetry around the times of eclipse. MeerKAT radio observations show a faint source which has a peak flux density of 30.7 $\pm$ 5.4 $\mu$Jy/beam at 1.28 GHz at the position of UZ For. 
\end{abstract}




\section{Introduction}

The AM Herculis (hereafter AM Her) systems, or polars, are a sub-class of magnetic cataclysmic variable (mCV) stars consisting of a strongly magnetized white dwarf primary ($B \approx$10--230 MG, e.g. \citealt{1996ApJ...473..483S}) and a low-mass main-sequence secondary star. 
The interaction between the magnetic field of the white dwarf (WD) and that of the mass-transferring secondary star results in synchronous rotation of the two stars \citep{1992Sci...258.1015F}. The presence of the strong magnetic field in the WD prevents the formation of an accretion disc. The red dwarf is constantly transferring material to the WD via Roche lobe overflow. 
Upon leaving the inner Lagrangian point ($L_1$), the material from the secondary star follows a ballistic stream trajectory and accelerates to supersonic speeds towards the WD until, at some distance from the WD, the magnetic pressure overwhelms the ram pressure of the ballistic stream. 
At this point, the ballistic stream is diverted from the orbital plane of the binary and follows a trajectory along the magnetic field lines of the WD before being accreted. For review on polars see, for example, \citet{1990SSRv...54..195C} or \citet{1995CAS....28.....W} or \citet{2001cvs..book.....H}. 

The material in the magnetic stream is ionized due to collisions within the stream and also by X-rays from the accretion region on the surface of the WD. 
At some height above the WD surface, a stand-off shock is formed when the supersonic free-falling material become subsonic. The typical temperature of the shock reaches 10--50 keV and this results in the gas being highly ionized.
The heated plasma cools as it settles onto the surface of the WD, resulting in a stratified post-shock region. 
In the post-shock region, the electrons and ions are forced to gyrate around the magnetic field lines, the former emitting cyclotron radiation which is beamed and highly polarized.
The AM Her types are recognized for their high degree of polarization (e.g. AM Her \citealt{1977ApJ...212L.125T}).
Cyclotron emission is thought to be caused by radiation from a hot plasma in a magnetic field of greater than 10 MG \citep{1981ApJ...244..569C,1982MNRAS.198...71M}. 
The post-shock region is also responsible for the emission of X-ray bremsstrahlung radiation and some of this is reprocessed by the surface of the WD and re-emitted as soft X-rays. 
Accretion onto the WD occurs over a small area on the surface of the WD, near one or both the magnetic poles. 

A number of mCVs studied in the optical led to the determination of their magnetic field strengths from cyclotron spectra \citep[see, e.g. ][]{1990MNRAS.243..565C,1992MNRAS.256..252F}.  
A common feature seems to be two-pole accretion, where the main accreting spot is at the lower field, e.g. VV Pup \citep{1989ApJ...342L..35W}, DP Leo \citep{1990MNRAS.245..760C}, WX LMi \citep{1999A&A...343..157R} and V1500 Cyg \citep{2018MNRAS.474.1572H}. 
Also, field strength in the main spot are always below 60 MG (except AR UMa with a probable field strength of 230 MG, \citealt{1996ApJ...473..483S}). 
A few other mCVs have been found to have the magnetic field strength higher than 60 MG from the main accreting spot, e.g. $\sim$90 MG for RX J1007.5-2016 \citep{1999ASPC..157..187R}, $\sim$150 MG for V884 Her \citep{2001ApJ...553..823S} and $\sim$110 MG for RX J1554.2+2721 \citep{2006A&A...452..955S}.

In recent studies, \cite{2018AJ....155...18L} used the homogeneous cyclotron-emission model of \cite{1979ApJ...232..895C} and \cite{1985MNRAS.214..605W} to constrain the magnetic field strength of the WD in MASTER J132104.04+560957.8 to be $\sim$30 MG. 
Most recently, \cite{2020MNRAS.491..201J} detected cyclotron harmonics in the optical spectra of three AM Her systems: RX J0859.1+0537, RX J0749.1--0549 and RX J0649.8--0737, which led to the determination of the magnetic field strength of the WD to be within $\sim$50 MG.

Doppler tomography is an indirect imaging technique that was developed by \cite{1988MNRAS.235..269M} which uses orbital phase-resolved spectra to construct a two-dimensional image in velocity space. 
This technique was developed to interpret emission-line profile variations of the accretion discs in non-magnetic CVs. 
Doppler tomography is governed by five axioms outlined in \cite{2001LNP...573....1M}, but violation of any of these axioms is possible. 
For example, \cite{2003MNRAS.344..448S} extended this technique by violating the second axiom which state \say{the flux of each element is constant} in order to isolate the emission components that vary with the spin or orbital period. \cite{2004MNRAS.348..316P}, for example, violated the first axiom when they applied the Doppler tomography technique to the spectra of V834 Cen since they only considered data covering half of the orbital phase. \cite{2015A&A...579A..77K} presented a complementary extension to the standard Doppler tomography technique called the inside-out projection. 
This method is more intuitive to interpret and redistributes the relative contrasts levels in and amongst the emission components. Also, Doppler maps of mCVs, specifically polars, have always been tricky to interpret since some of the motion is not confined to the orbital plane of the binary.   

The first radio detection of a mCV was reported by \citet{1982ApJ...255L.107C} for AM Her using the Very Large Array (VLA) at 5 GHz was found to have a flux density of 0.67 mJy. However, they did not detect any circular polarization despite the observations being taken in full polarization mode. 
Follow-up study by \cite{1983ApJ...273..249D} show a 100\% circularly polarized radio flare with a peak flux of 9.7 mJy lasting about ten minutes. The radio emission from AM Her was attributed to gyrosychrotron emission from energetic electrons trapped in the magnetosphere of the WD. The radio flare was attributed to an electron-cyclotron maser that operates near the surface of the red dwarf in a magnetic field of $\approx$1 kG. 
The second radio detection of a mCV was made by \cite{1988MNRAS.231..319W} for V834 Cen using the Parkes 64-m telescope at 8.4 GHz. They found V834 Cen to be variable on time-scales as short as one minute and reaching peak flux densities of 35 mJy. The emission from V834 Cen was attributed to an electron-cyclotron maser, although the maser emission suggestions in polars has recently been challenged by \citet{2019ARep...63...25K}. Instead they suggest the radio emission arises from Alfvenic turbulence.  

Several radio surveys of mCVs have been carried out in the past two decades by different authors.  
The survey of 22 previously unobserved mCVs by \citet{1994AJ....108.2207B} using both the VLA and the Australian Telescope Compact Array, at 8.4 GHz, yielded nondetections.  
In another survey of 21 mCVs made by \cite{1994MNRAS.269..779P} using the Jodrell Bank broadband interferometer, they detected five polars of which three were new detections: BG CMi, ST LMi and DQ Her. 
\cite{1996A&A...310..132M} observed BY Cam with the VLA telescope at three frequencies and their results only gave upper limits to the flux densities. 
The VLA observations at 8.4 GHz of nine mCVs within 100 pc by \cite{2007ApJ...660..662M} showed strong radio emission from AR UMa and AM Her.  
Recently, \cite{2017AJ....154..252B} detected radio emission from 18 mCVs from the survey of 121 mCVs observed with the Jansky Very Large Array (JVLA). Out of the 18 targets detected in the radio, 13 were new radio emitters and this increased the total number of known radio emitting mCVs to 21.  

UZ Fornacis (hereafter UZ For) is an eclipsing polar \citep{1987IAUC.4486....1G,1988ApJ...328L..45O} with an orbital period of 126.5 min and dM4.5 secondary star \citep{1988A&A...195L..15B}. 
Polarimetry studies of UZ For show polarization reaching about 10\% in circular and about 5\% in linear \citep{1988ApJ...329L..97B,1989ApJ...337..832F}. 
The spectra of UZ For has been presented in the literature \citep[see,][]{1988A&A...195L..15B,1989ApJ...337..832F,1989ApJ...347..426A}, and show hump features that were interpreted as due to cyclotron emission from a hot plasma \citep{1982MNRAS.198..975W}. 
Modelling of the cyclotron humps gave the first estimates of the magnetic field of UZ For to be $\sim$55 MG with the possibility of the second pole also emitting cyclotron radiation \citep{1988A&A...195L..15B,1989ApJ...347..426A}.  
A two-pole accretion model was invoked in order to explain the cyclotron humps with the main pole contributing the lower value of magnetic field \citep{1989ApJ...337..832F,1990A&A...230..120S}. 
\citet{1996A&A...310..526R} remodelled the observations from \cite{1990A&A...230..120S} by considering a WD heated by a stream of free falling electrons and they estimated a magnetic field strength of $\sim$53 MG and $\sim$48 MG for the two poles. 
\cite{2002ASPC..261..159N} found cyclotron harmonics in both the faint (low-accretion) and bright (high-accretion) state ultra-violet spectra of UZ For. They determined magnetic field strengths of 51 and 74 MG for the bright and faint phase, respectively.  
The distance to UZ For was estimated to be $\sim$240 pc by \citet{2019A&A...621A..31K} based on $GAIA$ parallax measurements. Prior to this study, there has not been any circular spectropolarimetry results presented for this target. 

In this paper we present optical and radio observations of UZ For. The structure of the paper is as follows. Section \ref{sect:obs} contains all the observations and reductions.  
Section \ref{sect:resul} contains our results and analysis. We provide a general discussion and conclusion in Sect. \ref{sec:dis}. 

\section{Observations}\label{sect:obs}

This section is structured as follows: Sect. \ref{sect:photom} explains the photometry observations taken with the 1.9-m telescope situated at the South African Astronomical Observatory (SAAO) in Sutherland; Sect. \ref{sect:spec} contains the spectroscopic observations taken with with the Southern African Large Telescope (SALT\footnote{More details on SALT can be found at \url{http://www.salt.ac.za}.}, \citealt{2006SPIE.6269E..0AB}); Sect. \ref{sect:polarimetry} discusses the polarimetry data obtained with the SAAO 1.9-m telescope, and Sect. \ref{sect:spectropol} discusses the spectropolarimetry taken with SALT. 
The last two sections, Sect. \ref{sect:radio_obs} and \ref{sect:meerlicht}, contains the radio and optical photometry observations taken with the MeerKAT radio telescope \citep{2018ApJ...856..180C,2016mks..confE...1J} located at the site of the South African Radio Astronomical Observatory (SARAO) and MeerLICHT \citep{2016SPIE.9906E..64B} telescope situated at SAAO site in Sutherland. 
The photometry, spectropolarimetry and radio observations were taken simultaneously, the first time these facilities have been used in such a manner.

\subsection{Photometry} \label{sect:photom}

\begin{figure*}
\begin{center}
\includegraphics[width = \textwidth ]{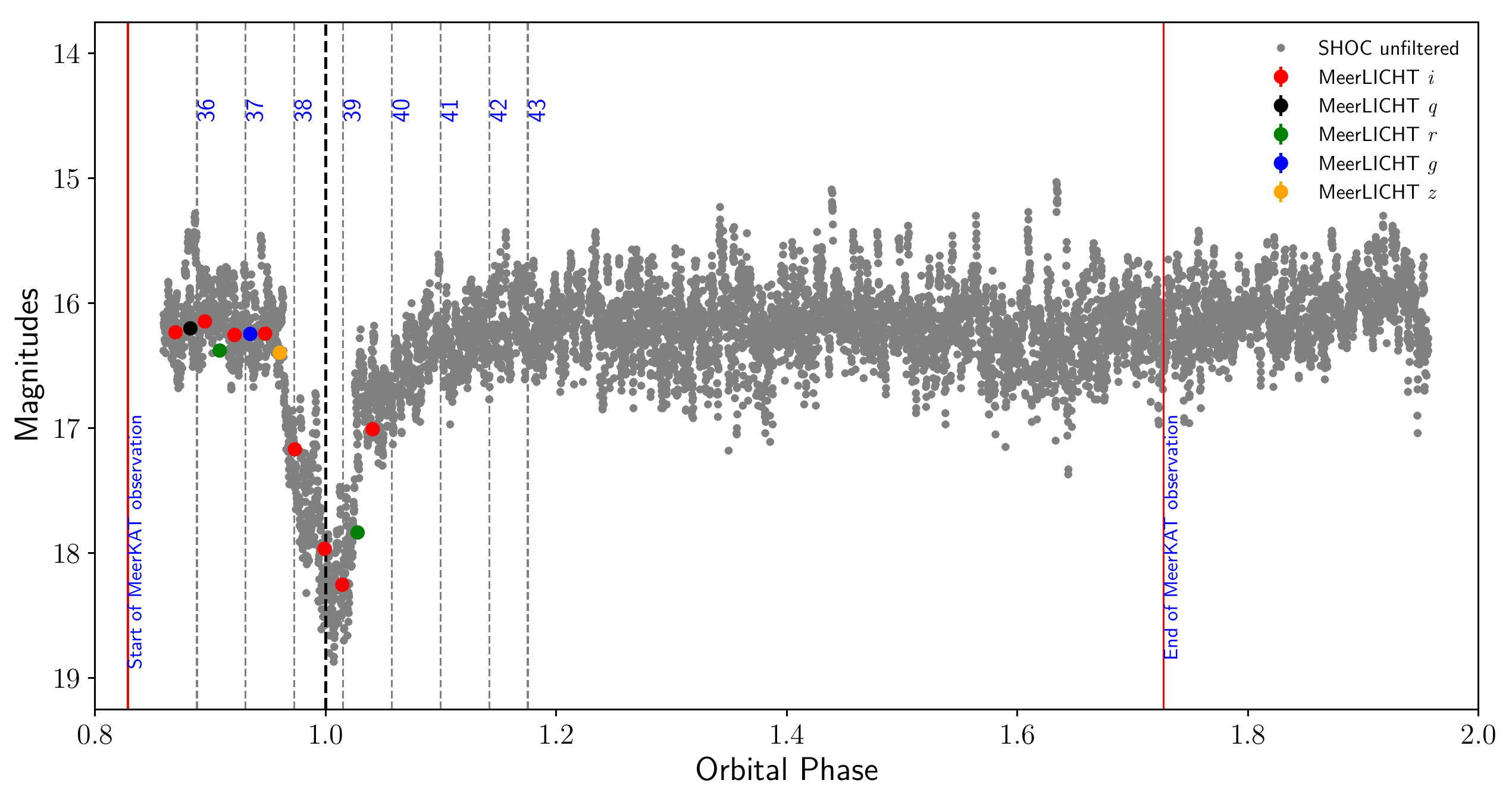}  
\end{center}
\caption{Simultaneous SHOC (grey dots) and MeerLICHT (filled circles) light curve of UZ For obtained on 2018 November 6, overlapping with SALT spectropolarimetric and MeerKAT radio observations. The vertical grey dotted lines mark the mid-exposure of the eight individual spectropolarimetry exposures taken with SALT whereas the black dashed line represent the time of mid-eclipse. The red vertical lines marks the start- and end-time of the MeerKAT observations.}
\label{figure:shoc}
\end{figure*}

High-speed photometric observations were obtained on the night of 2018 November 6 using the SAAO 1.9-m telescope that is equipped with the Sutherland High-speed Optical Camera (SHOC, \citealt{2011epsc.conf.1173G,2013PASP..125..976C}) in photometric conditions. 
The SHOC detector (an EM-CCD based system using an Andor iXon 888 camera) was used in frame-transfer mode with a clear filter and binning of 4 $\times$ 4.  
A cadence of one second was used, and the resulting data cubes were reduced using the SHOC pipeline that is described in \citet{2013PASP..125..976C}. 
The white light counts were converted to magnitudes by comparing the magnitudes of UZ For with the B and R magnitudes of stars in the field around UZ For and listed on the USNO catalogue \citep{2003AJ....125..984M}.
We corrected all the times for the light travel-time to the barycentre of the solar system (i.e. converted to the barycentric dynamical times (TDB) as Barycentric Julian Dates (BJD; \citealt{2010PASP..122..935E})). 

We used the ephemeris from \cite{2019A&A...621A..31K} to phase the lightcurve. The resulting lightcurve obtained after the data reduction is shown in Fig. \ref{figure:shoc} and discussed in Sect. \ref{sect:photo_lc}.  

\subsection{Spectroscopy}\label{sect:spec}

Spectroscopic observations of UZ For were made with SALT over five nights between 2013 January 03 and 2014 January 30 using the Robert Stobbie Spectrograph (RSS; \citealt{2003SPIE.4841.1463B,2003SPIE.4841.1634K}) in long-slit mode. Table \ref{tab:03} provides the observation log. A long-slit of width 1.5$''$ was used and at least seven medium resolution spectra of UZ For, each with exposure time of 360 s, were obtained per night. Two of the five observations were taken at longer (red) wavelengths and the remaining three at shorter (blue) wavelengths. 
For the blue spectra, the PG2300 grating was used at a grating angle of 32$^{\circ}$ and a camera station angle of 64$^{\circ}$. This gives a central wavelength of 4600\AA{} and a coverage of $\sim$4050--5100 \AA{} at a resolving power of $\approx$2300. For the red spectra, the PG1800 grating was used at a grating angle of 48$^{\circ}$ and a camera station angle of $\sim$95$^{\circ}$. This gives a central wavelength of 8600 \AA{} and a coverage of $\sim$7550--8650 \AA{} at a resolving power of $\approx$4000. 
These observations were timed such that they cover one orbital period of the binary. ThAr lamp exposures were taken at the end of each observation for the purpose of wavelength calibrations. 
In addition, the spectra of spectrophotometric standard stars (LTT 4364 and EG21) were obtained either on each night or a day(s) after the actual observations were taken for relative flux calibration. 

\begin{table*}
\caption{Spectroscopic, photometric, photopolarimetric, spectropolarimetric and radio observation log of UZ For.}
\label{tab:03}
\centering
\begin{tabular}{l c c c c c c c} \hline \hline
Date of     & Number of & Exposure or & Spectral & Wavelength & Type of &Telescope & Instrument \\
observation &  spectra  & integration &resolution& range   & observation & used    & Used \\
            & or points &  time (s)   & (mm)     & (\AA{}) & &  &  \\ \hline \hline
2013/01/03  &    7   &  360    &  4000 & 7550-8650  & spectroscopy & SALT & RSS \\
2013/01/06  &    7   &  360    &  2300 & 4050-5100  & spectroscopy & SALT & RSS \\
2013/01/07  &    8   &  360    &  2300 & 4050-5100  & spectroscopy & SALT & RSS \\
2013/01/08  &    8   &  360    &  2300 & 4050-5100  & spectroscopy & SALT & RSS \\
2014/01/30  &    7   &  360    &  4000 & 7550-8650  & spectroscopy & SALT & RSS  \\
2018/10/04  &  5434  &   1     &   -   &clear filter& photopolarimetry  & SAAO 1.9-m & HIPPO\\
2018/11/06  &   8    &   360   & 2500  & 3200-9000 & spectropolarimetry & SALT & RSS \\
2018/11/06  & 8330   &    1    &   -   & unfiltered & photometry & SAAO 1.9-m & SHOC \\
2018/11/06  &  14    &    60    &   -   & $i,q,r,g,z,u$ & photometry & MeerLICHT & STA1600 \\
2018/11/06  & -      &    8    &   -   &    -     & radio imaging & MeerKAT & L-band  \\ \hline 
\end{tabular}
\end{table*}

Data reduction was carried out using the \textsc{pysalt} software package\footnote{For more details on pysalt visit \url{http://pysalt.salt.ac.za/}.} \citep{2010SPIE.7737E..25C} and \textsc{iraf}\footnote{IRAF is distributed by the National Optical Astronomy Observatories,  which  are operated by  the Association of Universities for Research in Astronomy, Inc., under cooperative agreement with the NSF.} reduction procedures. These included overscan correction, bias subtraction and gain correction. 
Relative flux correction of the blue and red spectra were based on the sensitivity of the spectrophotometric standard stars LTT 4364 and EG21 \citep{1984MNRAS.206..241B}, respectively. 
Extinction correction was applied to the resulting spectra of UZ For and Doppler corrections due to the motion of the Earth around the Sun were removed.  
The average blue and red spectra of UZ For are presented in Fig. \ref{figure:specBR}. The two gaps in the spectra are due to the RSS detector consisting of a mosaic of three chips, and this in turn results in two small gaps in the wavelength dispersion direction. The width of the gaps is $\sim$10 \AA{}.

\subsection{Photopolarimetry}\label{sect:polarimetry}

Photopolarimetry observations of UZ For were made on the night of 2018 October 4 with the SAAO 1.9-m telescope using the HIgh-speed Photo-POlarimeter (HIPPO, \citealt{2010MNRAS.402.1161P}). The HIPPO instrument was operated in its simultaneous linear and circular polarimetry and photo-polarimetry mode (all-Stokes). 
The observations were clear filtered, defined by the response of the two RCA31034A GaAs photomultiplier tubes which give the wavelength coverage from 3500-9000\AA{}.   
Polarized and non-polarized standard stars \citep{1982ApJ...262..732H,1988AJ.....95..900B} were observed on the night in order to calculate the waveplate position angle offsets, instrumental polarization, and efficiency factors.  
Background sky measurements were taken at frequent intervals during the observations. No photometric calibrations were carried out; photometry is given as the total counts minus the background-sky counts (taken from the preceding sky observation). All of our observations were synchronized to a GPS clock to better than a millisecond. As with the SHOC observations, we corrected all the times to the barycentre of the solar system. 
Data reduction was carried out following the procedures described in \citet{2010MNRAS.402.1161P}. The resulting photometry light curve, percentage circular and linear polarization are shown in Fig. \ref{figure:pol_lc} and discussed in Sect. \ref{sect:polar}. We calculated the phase using the ephemeris derived in \cite{2019A&A...621A..31K}. 

\subsection{Spectropolarimetry}\label{sect:spectropol}

UZ For was observed with the SALT RSS instrument in circular spectropolarimetry mode \citep{2003SPIE.4843..170N} on the night of 2018 November 6. The spectropolarimetry uses a rotating quarter- and half-waveplates near the focal plane and a calcite mosaic beamsplitter before the camera. 
A total of eight exposures of 300 s each, with orbital phase resolution of about 0.05 and spectral resolution of 4 \AA{}, containing the ordinary (O) and extraordinary (E) beams, were obtained around the eclipse (see Fig. \ref{figure:shoc}). 
An exposure of Argon lamp was taken after the science frames for wavelength calibration purposes. We used the PG0300 grating providing a resolving power of $\sim$2500 and a wavelength coverage of $\sim$3200--9000 \AA{}.   
The observations of the spectrophotometric standard star (HILT600, unpolarized optical calibrator) were obtained on the night of 2018 December 4 with the same setup as our science exposures. 

The CCD pre-processing of the observations was performed using the polsalt-beta\footnote{See \url{https://github.com/saltastro/polsalt/} for more details.} software \citep{2003SPIE.4843..170N,2012AIPC.1429..248N,2016SPIE.9908E..2KP} based on \textsc{pysalt} package \citep{2010SPIE.7737E..25C}, this includes overscan correction, bias subtraction and gain correction. 
The wavelength calibration for both the O and E beams was performed using the Argon lamps taken with the same observation septup. The E and O beams of each spectra were extracted using the polsalt software and stored as two different extensions in a single output file. These extensions, containing the one dimensional extracted O and E beam spectra, were then split into two separate files via a script we wrote in \textsc{python}. 

We computed the degree of circular polarization ($V/I$) from two consecutive exposures, with the quarter wave retarder plate rotated by $\pm$45$^{\circ}$, using Equ. \ref{equ:vi} below (adopted from \citealt{2005A&A...442..651E}):   

\begin{equation}
\label{equ:vi}
\frac{V}{I} = \frac{1}{2} \left[ \left( \frac{f^{\rm{o}} - f^{\rm{e}}}{f^{\rm{o}} + f^{\rm{e}}}\right)_{\theta = 315^{\circ}} - \left( \frac{f^{\rm{o}} - f^{\rm{e}}}{f^{\rm{o}} + f^{\rm{e}}}\right)_{\theta = 45^{\circ}}  \right] , 
\end{equation}
\noindent
where 45$^{\circ}$ and 315$^{\circ}$ indicate the position angle of the quarter wave plate and $f^{\rm{o}}$ and $f^{\rm{e}}$ are the ordinary and extraordinary beams, respectively. 
The total relative intensity was obtained by adding the sum of the O and E beams. For relative flux calibration we used the spectrophotometric standard star HILT600. 
The resulting total relative flux, percentage circular polarization and total polarized flux spectra are shown in Fig. \ref{figure:circ1} and discussed in Section \ref{sect:spectro}. The spectra were not corrected for telluric bands.  

\subsection{MeerKAT radio observation}\label{sect:radio_obs}

Observations of UZ For and the surrounding field were taken on 2018 November 6 (MJD 58428) using the MeerKAT radio telescope \citep{2018ApJ...856..180C,2016mks..confE...1J}.  
MeerKAT has a field of view of one square degree at 1.4 GHz. These observations were part of the ThunderKAT (The Hunt for Dynamic and Explosive Radio Transients with MeerKAT) Large Survey Project \citep{2017arXiv171104132F}. 
The observations were taken using 62 of the MeerKAT antennas, at a central frequency of 1.28\,GHz, with a total bandwidth of 856\,MHz split into 4096 channels. Observations started at 20:06:17.7 (Universal Time Central, \textsc{utc}) and finished at 21:59:58.9 (\textsc{utc}), overlapping with both the photometric and spectropolarimetric observation taken in Sutherland.  
Visibilities were recorded every 8 seconds. The band-pass and flux calibrator, PKS J0408-6545, was observed for 10 minutes at the beginning of the observation. Thereafter the gain calibrator, PKS J0409-1757, and UZ For were observed, for  approximately 1.5 minutes and 15 minutes respectively, alternating between them repeatedly for the remainder of the observation. The total integration time on UZ For was approximately 100 minutes.

The data were flagged using AOFlagger\footnote{See \url{https://sourceforge.net/projects/aoflagger/} for more details.} (version 2.9.0, \citealt{2010ascl.soft10017O,2012A&A...539A..95O}), i.e. removing radio frequency interferences (RFI). 
After flagging, the raw data was binned into 8 channels per bin, resulting in 512 channels with a channel width of 1.67 MHz each. 
Data reduction and first generation calibration were executed using standard procedures in CASA\footnote{See \url{https://casa.nrao.edu/} for more details.} (version 4.7.1, \citealt{2007ASPC..376..127M}). 
We made use of the facet based radio-imaging package DDFacet \citep{2018A&A...611A..87T} for imaging, implementing the SSDClean deconvolution algorithm and Briggs weighting with a robust parameter of 0.0. No self-calibration was implemented. Fitting was done in the image domain using the IMFIT task in CASA and noise levels were measured in the vicinity of the expected position of the source. 

\subsection{MeerLICHT photometry}\label{sect:meerlicht}

\begin{table}
\caption{Multi-filter photometric magnitudes from MeerLICHT.}
\label{tab:meerlicht}
\begin{center}
\begin{tabular}{ccccc} \hline
Time in  & Magnitude & Magnitude& Filter &  MJD-OBS \\ 
 \textsc{utc} &      & error   & used & \\ \hline
20:11:31 &  16.2327  & 0.0164  & $i$  & 58428.8413380238 \\         
20:13:09 &  16.2014  & 0.006   & $q$  & 58428.8424708274 \\         
20:14:45 &  16.1457  & 0.0151  & $i$  & 58428.8435791517 \\         
20:16:22 &  16.3788  & 0.0117  & $r$  & 58428.8447083269 \\         
20:18:00 &  16.2537  & 0.0156  & $i$  & 58428.8458396201 \\         
20:19:43 &  16.2458  & 0.0083  & $g$  & 58428.8470326267 \\         
20:21:21 &  16.2433  & 0.0156  & $i$  & 58428.8481645506 \\         
20:22:58 &  16.3968  & 0.0394  & $z$  & 58428.8492861561 \\         
20:24:37 &  17.1695  & 0.0323  & $i$  & 58428.8504328602 \\  
20:26:15 &    -      &    -    & $u$  & 58428.85156998667 \\
20:27:53 &  17.9652  & 0.0581  & $i$  & 58428.8527001744 \\         
20:29:50 &  18.2533  & 0.0763  & $i$  & 58428.8540511102 \\         
20:31:30 &  17.8352  & 0.033   & $r$  & 58428.8552131268 \\         
20:33:09 &  17.008   & 0.0264  & $i$  & 58428.8563639454 \\ \hline
\end{tabular}
\end{center}
Notes: \textsc{utc} -- universal time central and MJD-OBS -- Modified Julian Date of the observations. The $u$ filter yielded no measurement.
\end{table}

We also obtained photometric data using the MeerLICHT\footnote{MeerLICHT is a prototype telescope for the BlackGEM telescope array; see \url{www.meerlicht.org}.} telescope \citep{2016SPIE.9906E..64B}. The MeerLICHT telescope is a fully robotic 0.65-m telescope with an instantaneous field of view matching that of MeerKAT. 
It is equipped with an STA1600 detector which provides a 2.7 square degree field of view at 0.56 arcsec/pixel. 
The observations started at 20:11 (UTC) and lasted for 22 minutes. Individual 60 s exposures in $g$, $r$, $z$, $q$ filters were interleaved with 60 s exposures in the $i$ filter. 
The MeerLICHT observations were taken during science commissioning of the telescope and covered most of the primary eclipse. 
Data were processed using the MeerLICHT pipeline which is a combination of tools from the Terapix software suite (e.g.  \citealt{1996A&AS..117..393B}) and the ZOGY image subtraction routines \citep{2016ApJ...830...27Z}, coded up by Paul Vreeswijk on behalf of the BlackGEM/MeerLICHT teams. 
Photometry for UZ For was extracted using the optimal photometry routines as outlined by \citet{1986PASP...98..609H} and \citet{1998MNRAS.296..339N}. The photometry was calibrated using a multi-mission, multi-wavelength all-sky photometric standard star database. Both the processing as well as the photometric calibration will be full discussed in a forthcoming paper. As with shoc data, the we converted the times from MJD to BJD and calculated the phases using the ephemeris from \cite{2019A&A...621A..31K}. 
The resulting lightcurve is shown in Fig. \ref{figure:shoc}. Table \ref{tab:meerlicht} shows the resulting magnitudes. 


\section{RESULTS AND ANALYSIS}\label{sect:resul}

\subsection{Photometry}\label{sect:photo_lc}

Figure \ref{figure:shoc} shows the light curve of UZ For obtained with the SHOC instrument. Overlaid on the plot are MeerLICHT exposures taken with $i, q, r, g$ and $z$ filters around the eclipse. The variation in magnitudes from MeerLICHT traces the primary eclipse of the binary system. 
The approximated magnitudes from SHOC, calculated based on the $B$ and $R$ magnitudes of the stars from the USNO catalogue and in the field around UZ For, agree with those from MeerLICHT telescope. 
The photometric observations were taken simultaneously with the SALT spectropolarimetry and MeerKAT radio observations. 
The vertical dotted grey lines marks the position of the mid-exposure of the spectropolarimetric observations labeled with the numbers 36 to 43. The vertical black dashed line indicates the position of mid-eclipse of the WD. The light curve of UZ For shows a lot of flickering when the system is out of eclipse. 
The shape of the eclipse is similar to those recorded in the literature (e.g. \citealt{1991MNRAS.253...27B,2019A&A...621A..31K}) and the out-of-eclipse shape of the light curve is similar to that of \cite{2001MNRAS.324..899P}.   

\subsection{Spectroscopy and Doppler tomography}
\label{sect:spec_an}

\begin{figure*}
\centering
\includegraphics[width = 0.9\textwidth  ]{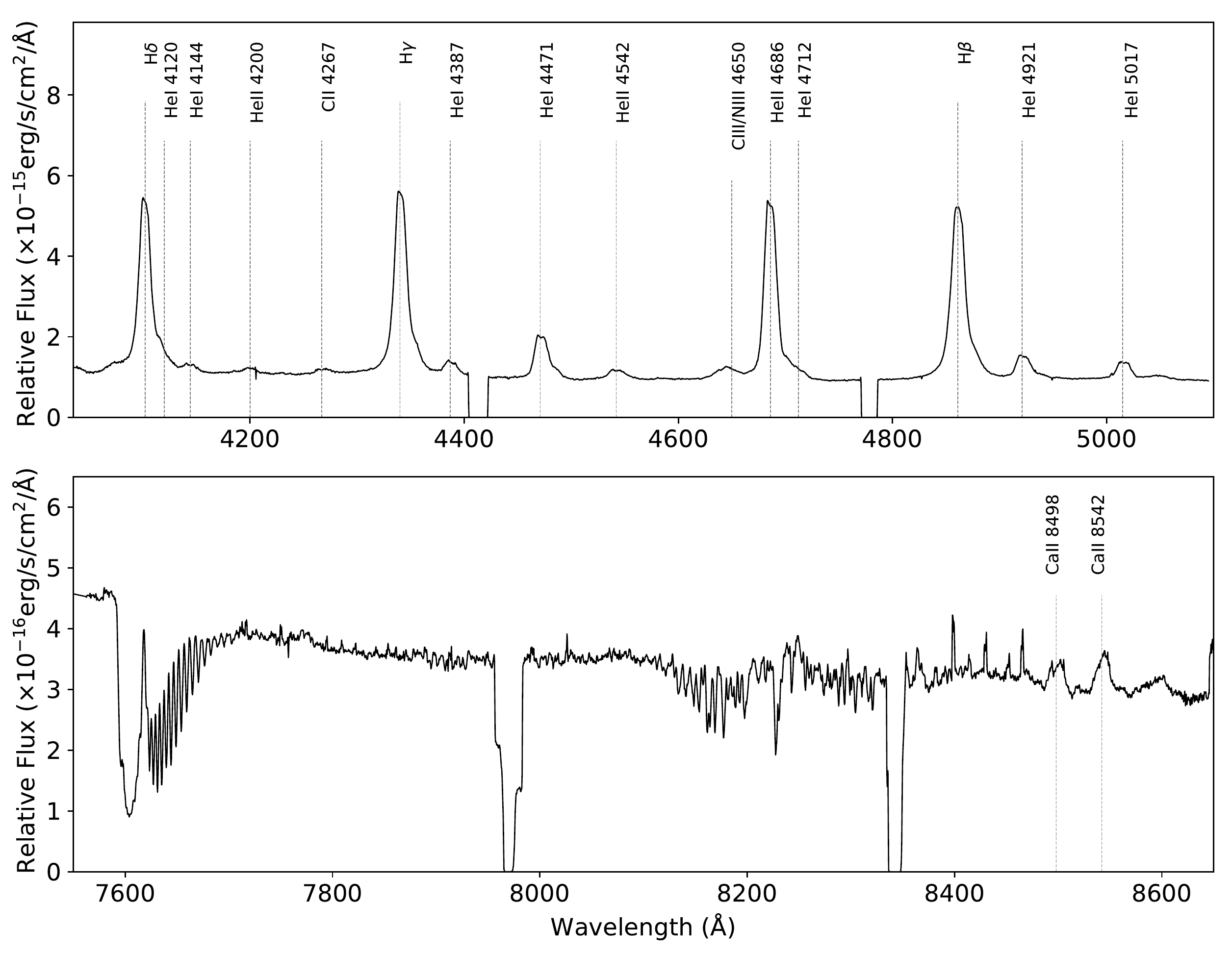}
\caption{Averaged wavelength calibrated blue (top) and red (bottom) spectra of UZ For obtained with SALT. Prominent emission and absorption features have been labeled.}
\label{figure:specBR}
\end{figure*}

Figure \ref{figure:specBR} (top panel) shows the averaged blue spectrum of UZ For taken over three nights. 
The blue spectrum shows the presence of single- and/or double-peaked emission from the Balmer lines, HeII lines (at 4200, 4542, 4686 \AA{}) and HeI lines (at 4120, 4144, 4367, 4471, 4713, 4921 and 5017 \AA{}). 
HeII 4686\AA{} and the Balmer lines dominate the continuum. The Bowen (CIII/NIII) blend at 4650\AA{} and CII 4267\AA{} are also present and appear in weak emission. 
The red spectrum of UZ For, shown in the bottom panel of Fig. \ref{figure:specBR}, shows a continuum dominated by telluric lines in absorption. 
A band at $\sim$7600 \AA{} is attributed to absorption by the Earth's atmosphere. There is strong emission from the CaII lines at 8498 \AA{} and 8542 \AA{}, likely from the irradiated secondary star. 


We used the strongest features from the blue spectra to compute Doppler maps of emission lines for further investigation utilizing the Doppler tomography code\footnote{See \url{http://www.saao.ac.za/~ejk/doptomog/main.html} for more details.}, described in \cite{2015A&A...579A..77K}. 
The inside-out method reverses the standard velocity projection by transposing the zero-velocity origin to the outer circumference of the map and uses polar coordinates. 
Also, it offers better spatial resolution to the higher velocity material which are stretched in the standard Doppler tomography techniques. 
This method has been applied to mCVs, (e.g.  \citealt{2016A&A...595A..47K,2017A&A...608A..36T,2019ApJ...881..141L}). 
We investigated all the Doppler maps of the strongest features: HeII 4686\AA{} and Balmer lines in the blue as well as CaII lines in the red. Here we are only presenting the Doppler maps of the H$\beta$, HeII 4686 \AA{} and CaII 8542\AA{} lines. 
The Doppler maps and trailed spectra presented here were normalized by the maximum flux in the input spectra for each spectrum.

\begin{figure}
\begin{center}$
\begin{array}{cc}
\includegraphics[height= 4cm, width=4cm]{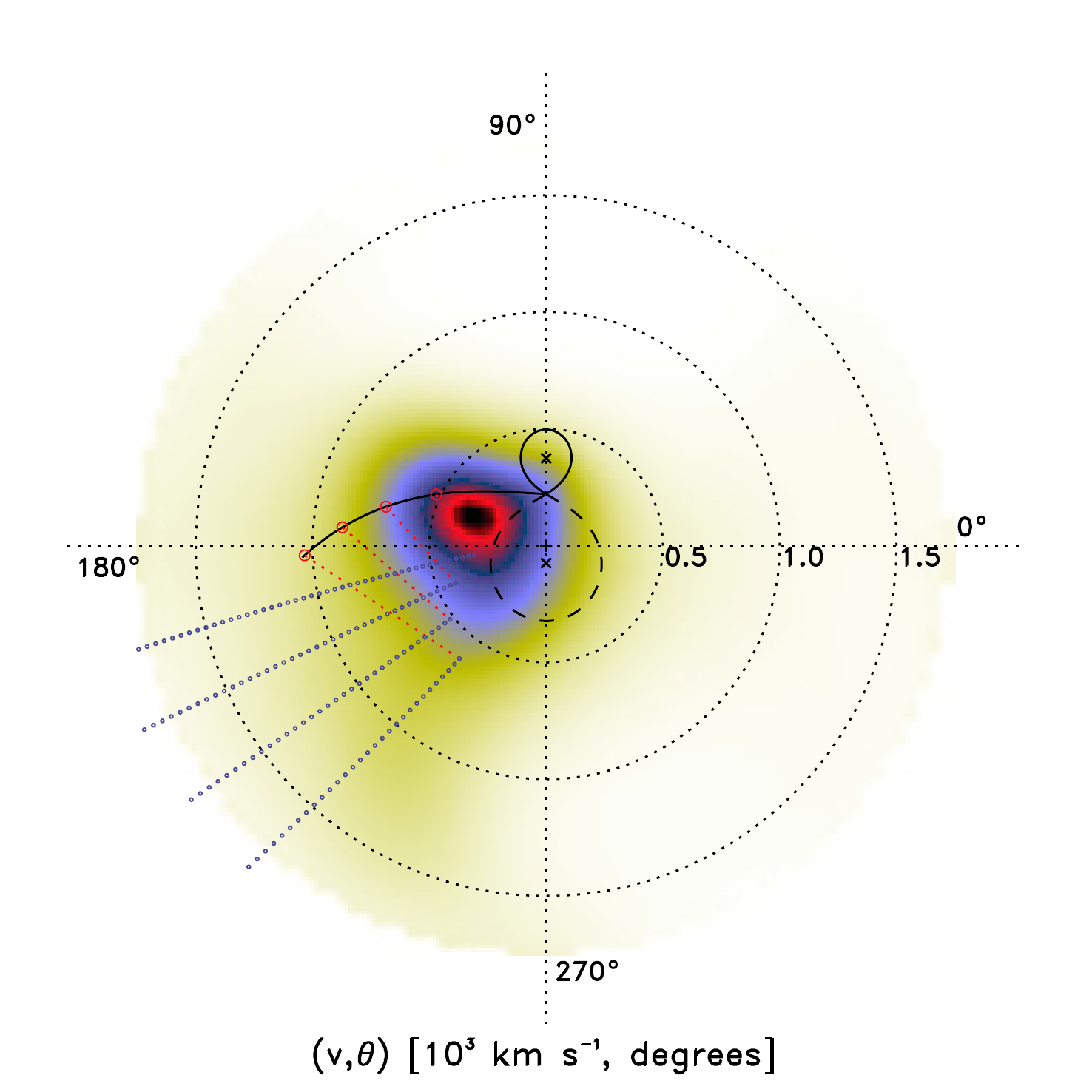} 
 \hspace{-0.25cm}
\includegraphics[height= 4cm, width=4cm]{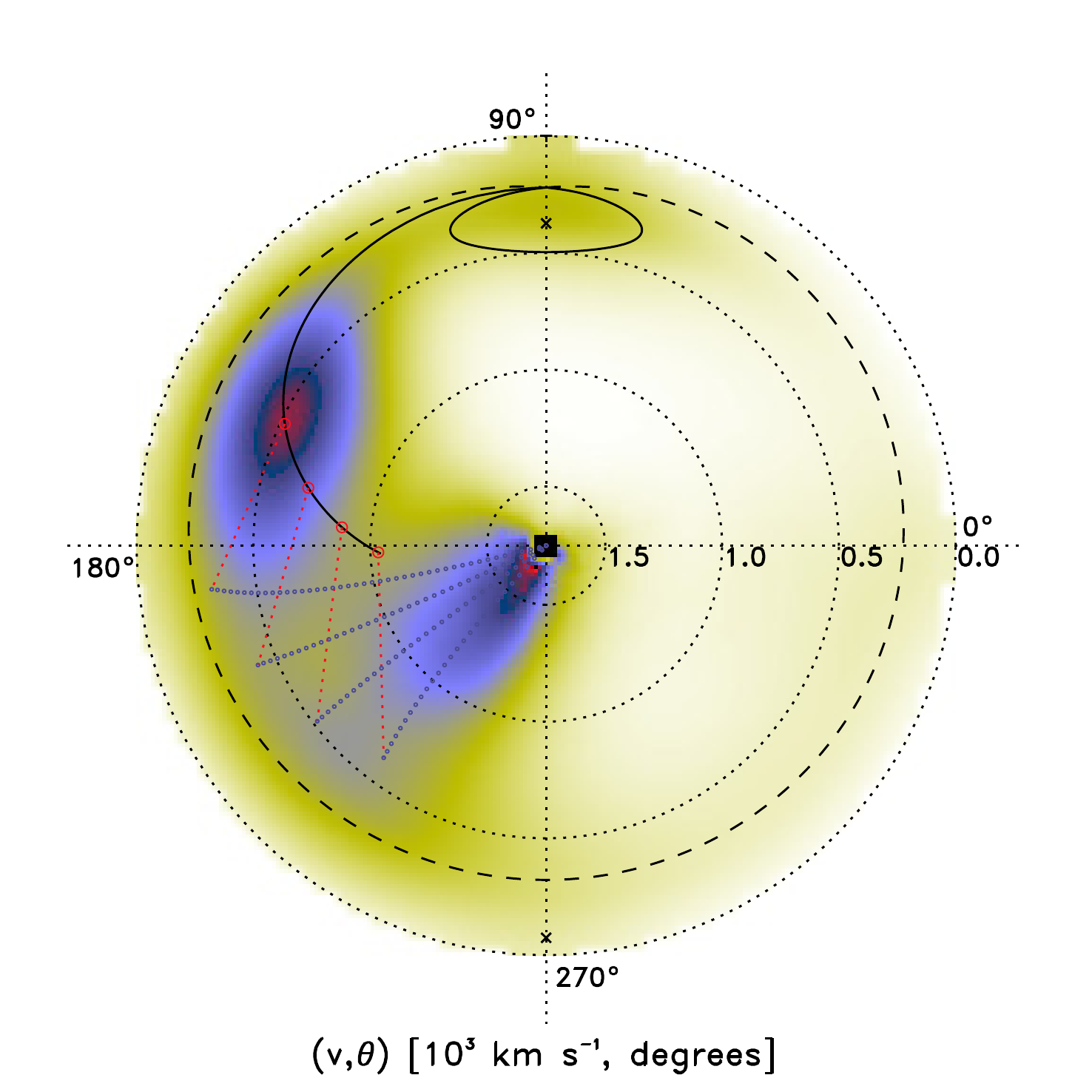}
\end{array}$
\end{center}
\vspace{-0.15cm}
\begin{center}$
\begin{array}{cccc}
\includegraphics[width=0.15\textwidth, height= 4.5cm]{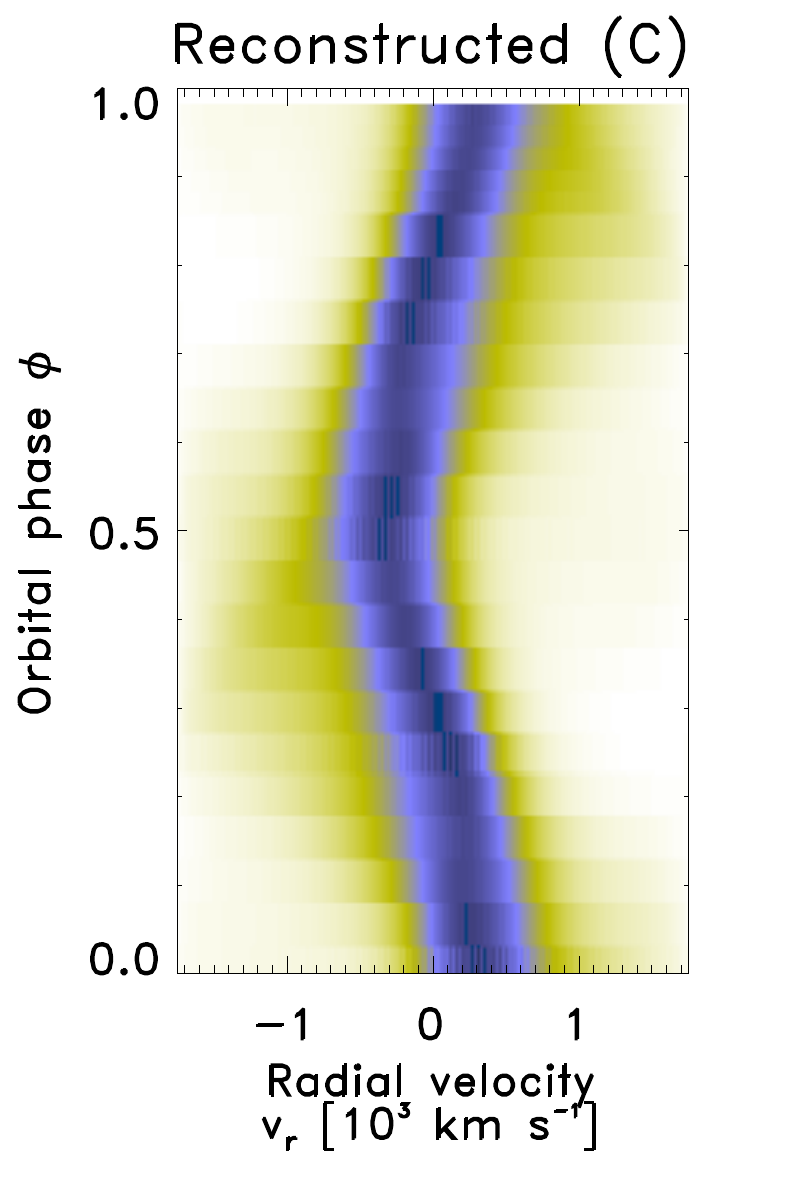} 
 \hspace{-.10cm}
\includegraphics[width=0.15\textwidth, height= 4.5cm]{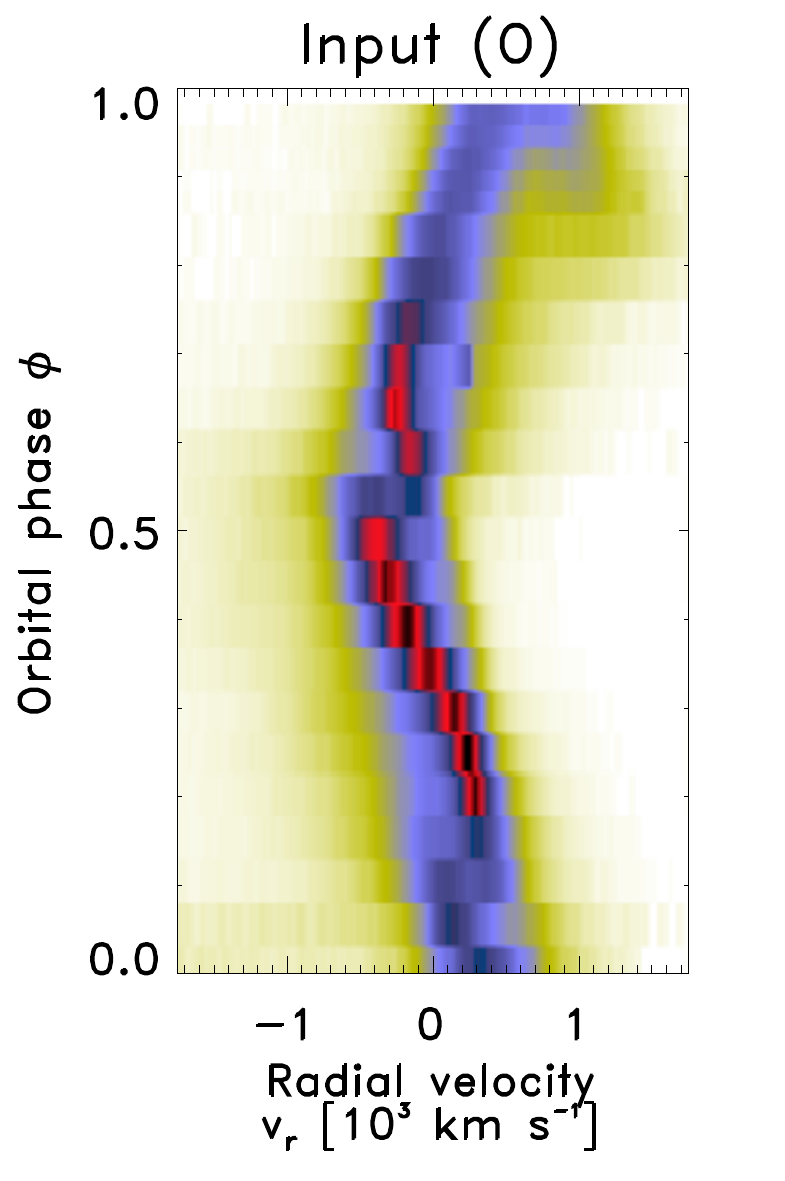} 
 \hspace{-.10cm}
\includegraphics[width=0.15\textwidth, height= 4.5cm]{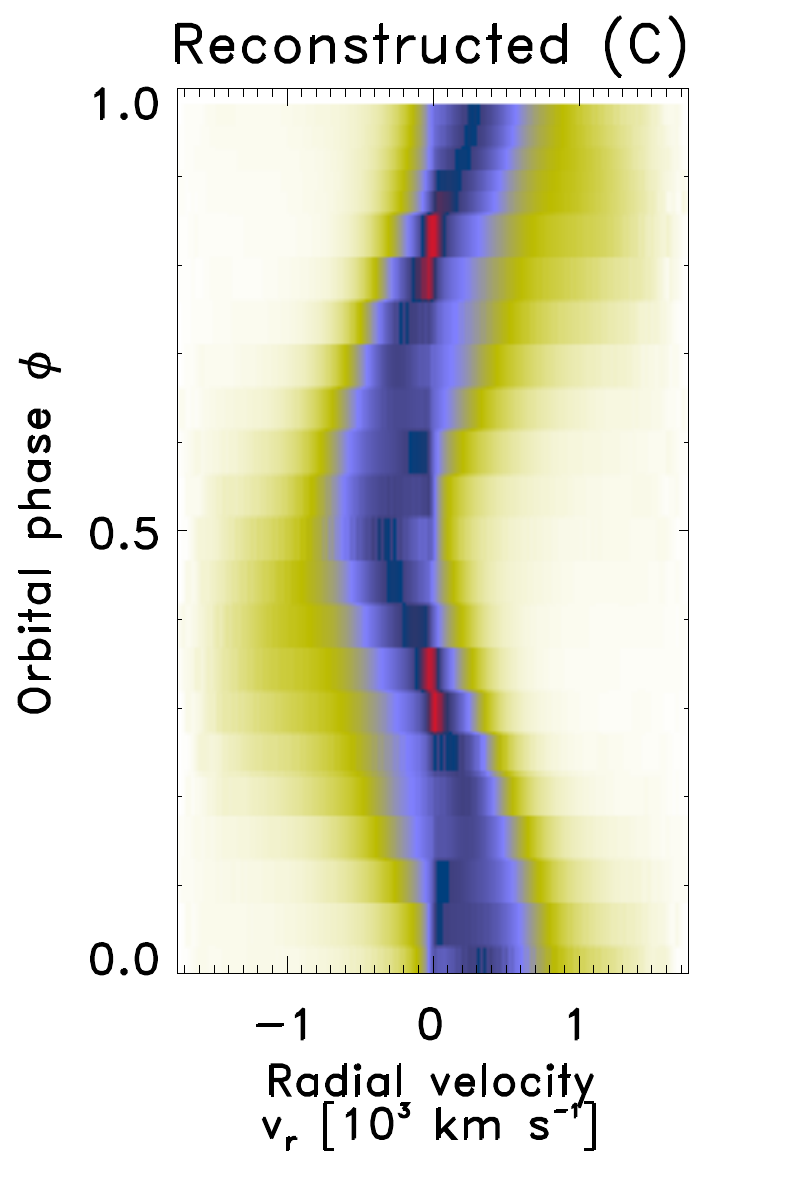}
 \hspace{-0.30cm}
\includegraphics[width=1.cm, height= 4.50cm]{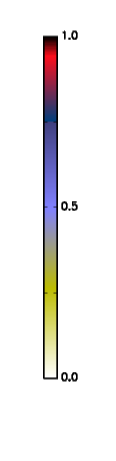} \\ 
\end{array}$
\end{center}
\caption{Standard and inside-out Doppler tomograms as well as trailed observed and reconstructed spectra based on the H$\beta$ emission line. Top row: the standard (left) and inside-out (right) Doppler tomograms. Second row: the input trailed spectra (centre) with the reconstructed trailed spectra for the standard (left) and inside-out (right) tomograms, respectively.}
\label{figure:doppsBeta_A}
\end{figure}

\begin{figure}
\begin{center}$
\begin{array}{ccc}
\includegraphics[height= 4.cm, width=4.cm]{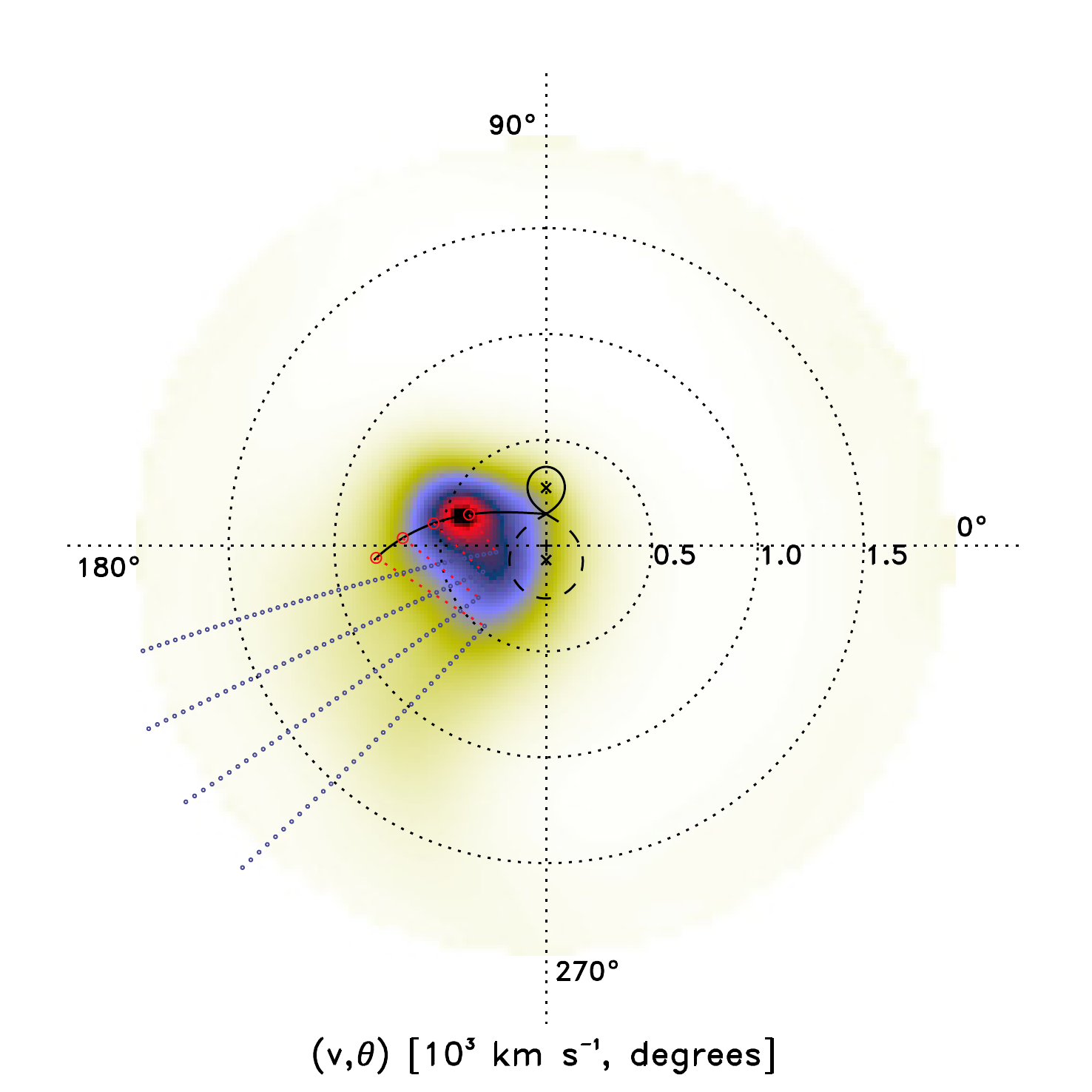} 
 \hspace{-0.25cm}
\includegraphics[height= 4.cm, width=4.cm]{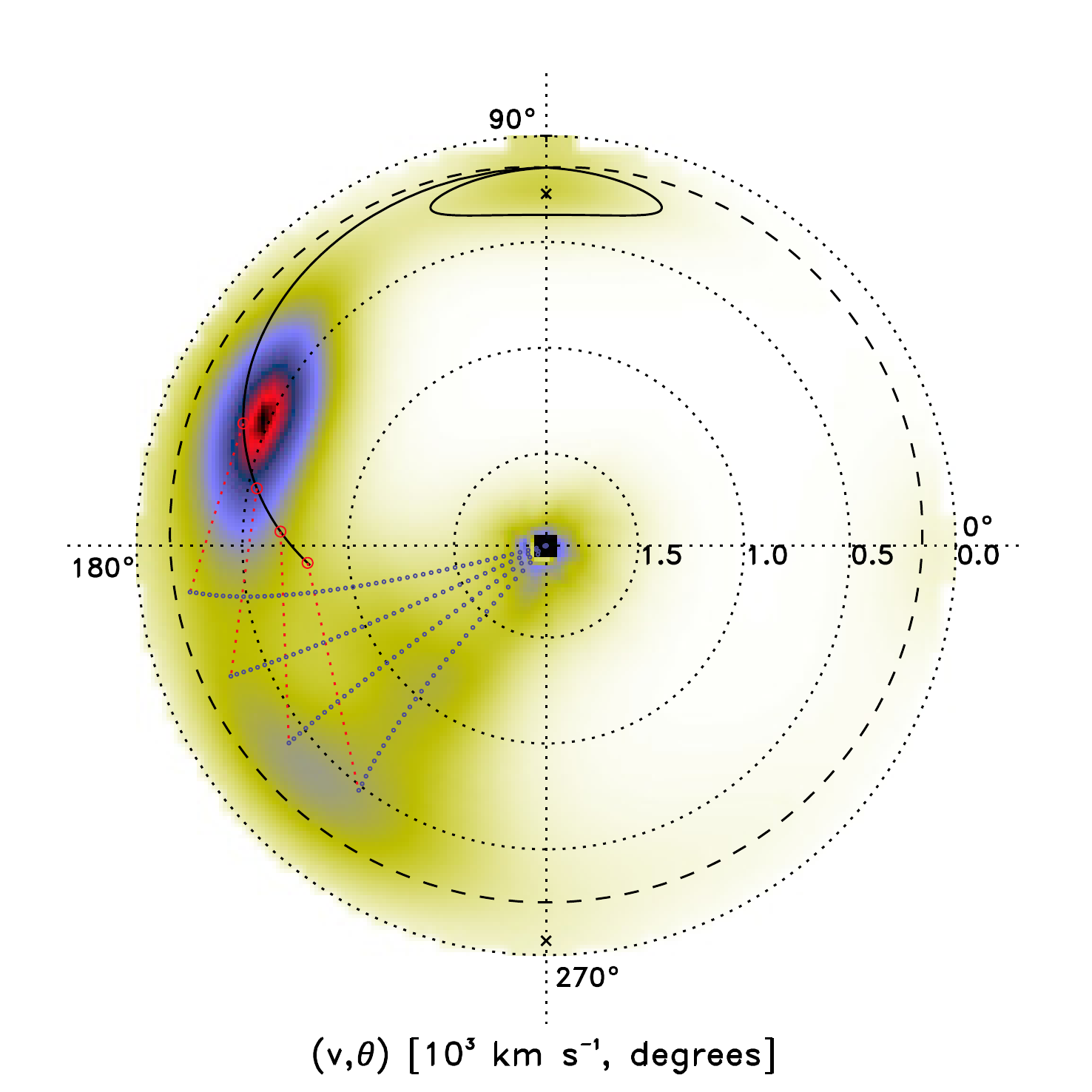}
\end{array}$
\end{center}
\vspace{-0.15cm}
\begin{center}$
\begin{array}{cccc}
\includegraphics[width=0.15\textwidth, height= 4.5cm]{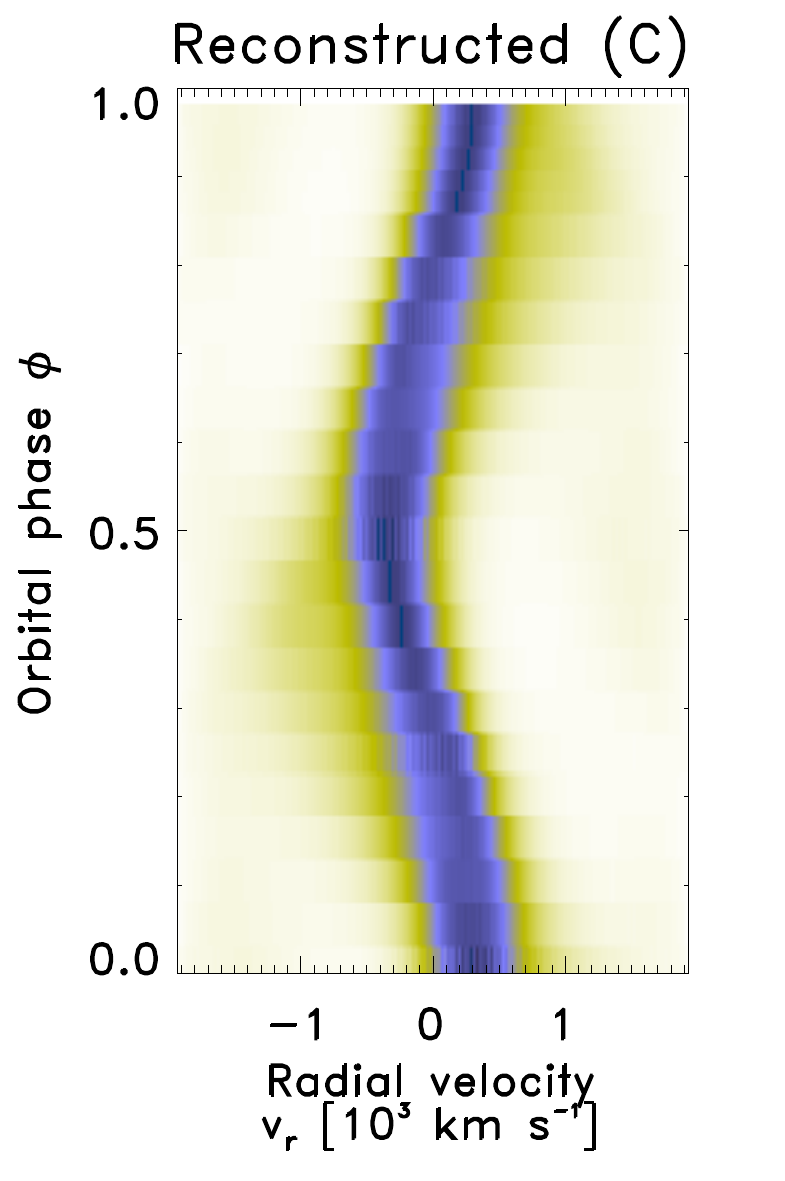} 
 \hspace{-.10cm}
\includegraphics[width=0.15\textwidth, height= 4.5cm]{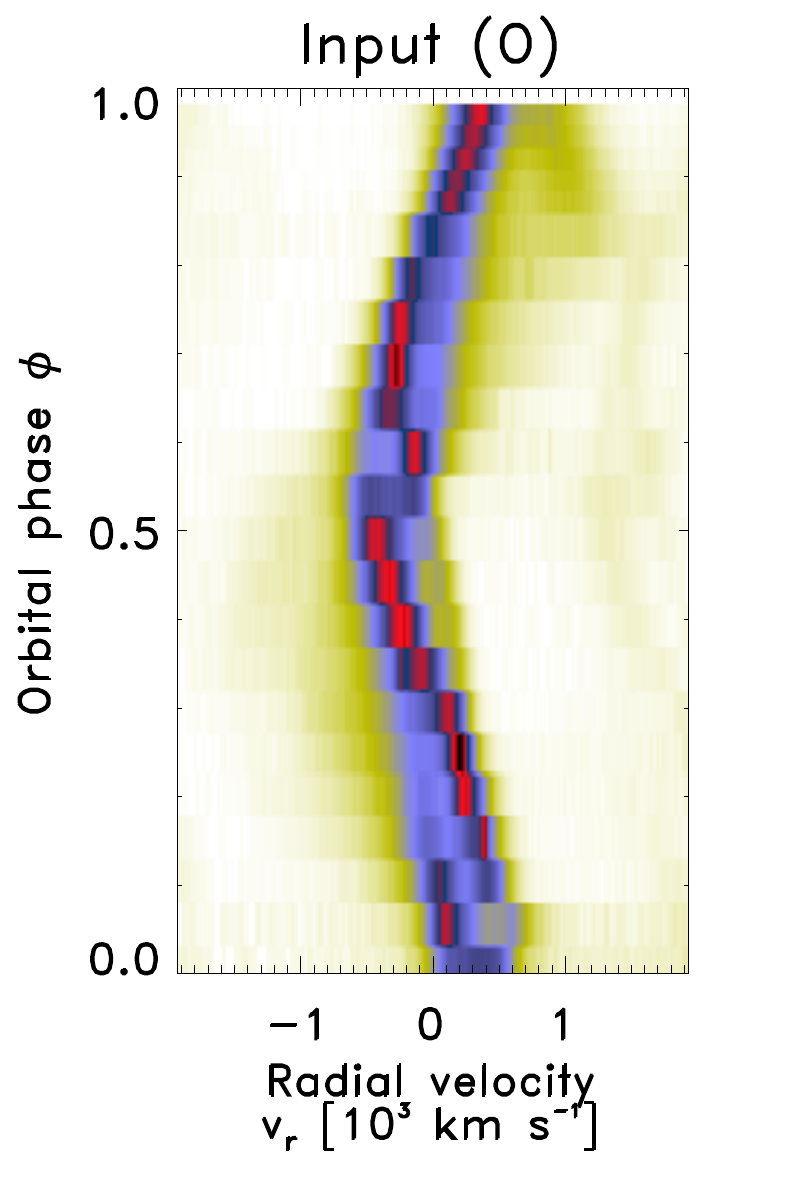} 
 \hspace{-.10cm}
\includegraphics[width=0.15\textwidth, height= 4.5cm]{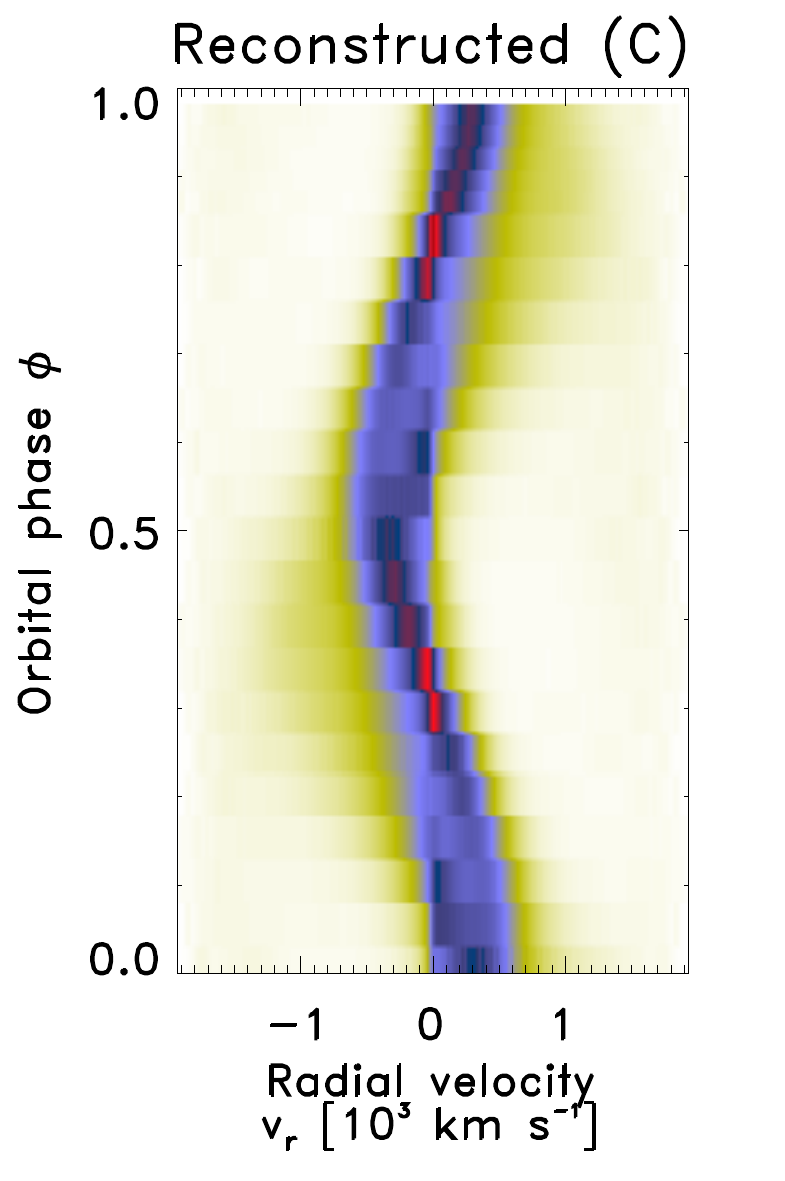}
 \hspace{-.30cm}
\includegraphics[width=1.cm, height= 4.5cm]{Spectra_colorbar_20.png} \\ 
\end{array}$
\end{center}
\caption{Same as Fig. \ref{figure:doppsBeta_A} but for HeII 4686 \AA{}.}
\label{figure:doppsHeII_A}
\end{figure}

\begin{figure}
\begin{center}$
\begin{array}{ccc}
\includegraphics[height= 4.0cm, width=4.cm]{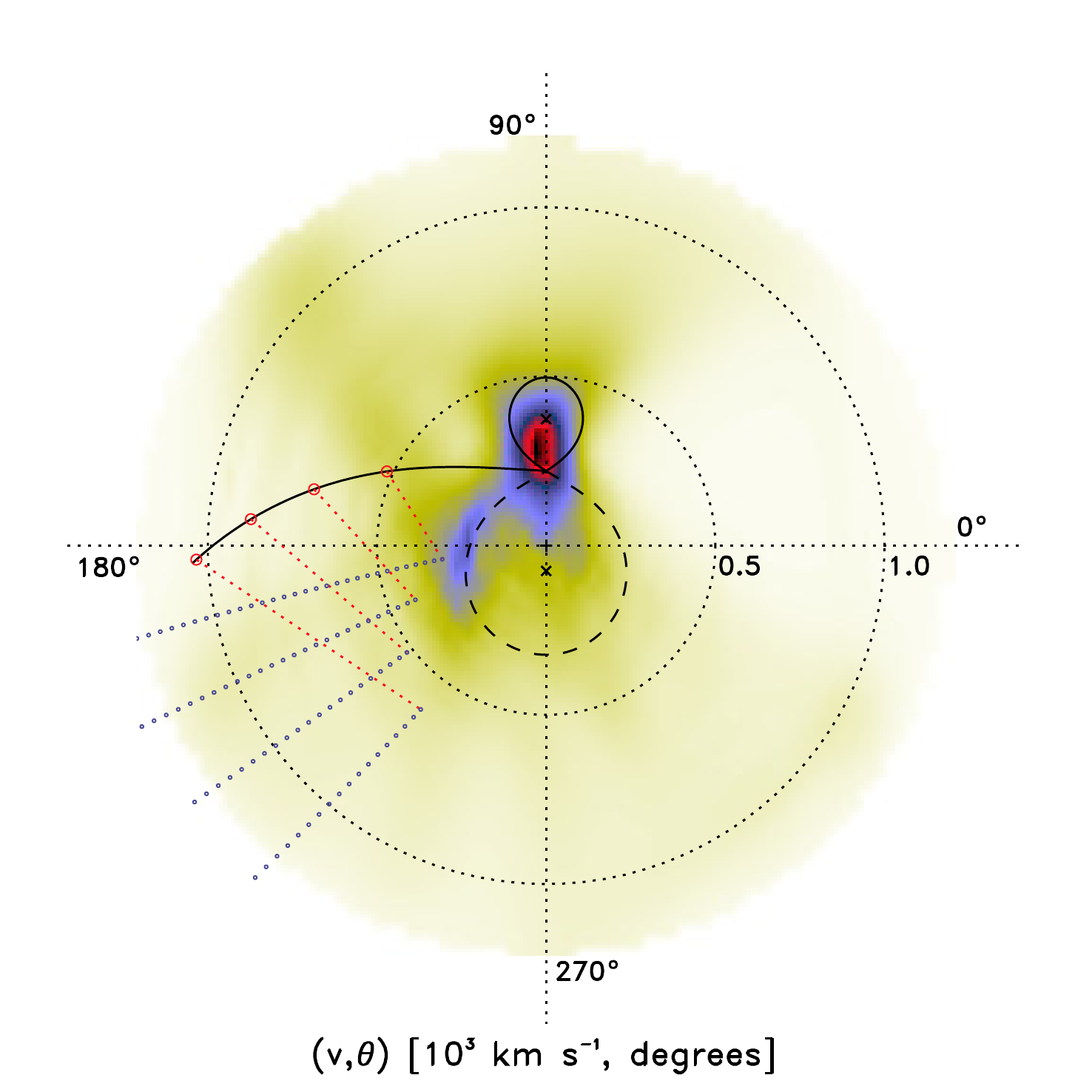} 
 \hspace{-0.25cm}
\includegraphics[height= 4.0cm, width=4.cm]{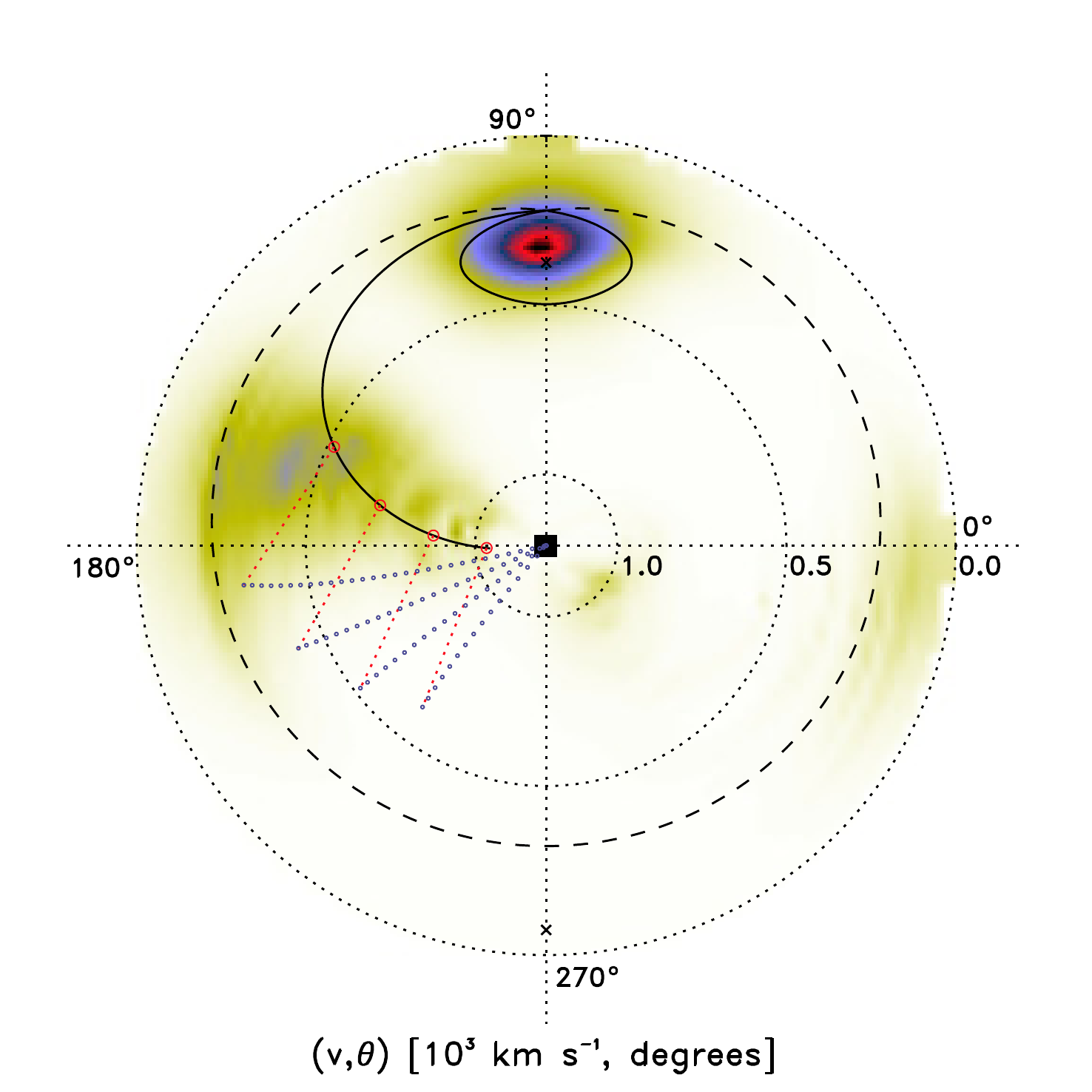}
\end{array}$
\end{center}
\vspace{-0.15cm}
\begin{center}$
\begin{array}{cccc}
\includegraphics[width=0.15\textwidth, height= 4.5cm]{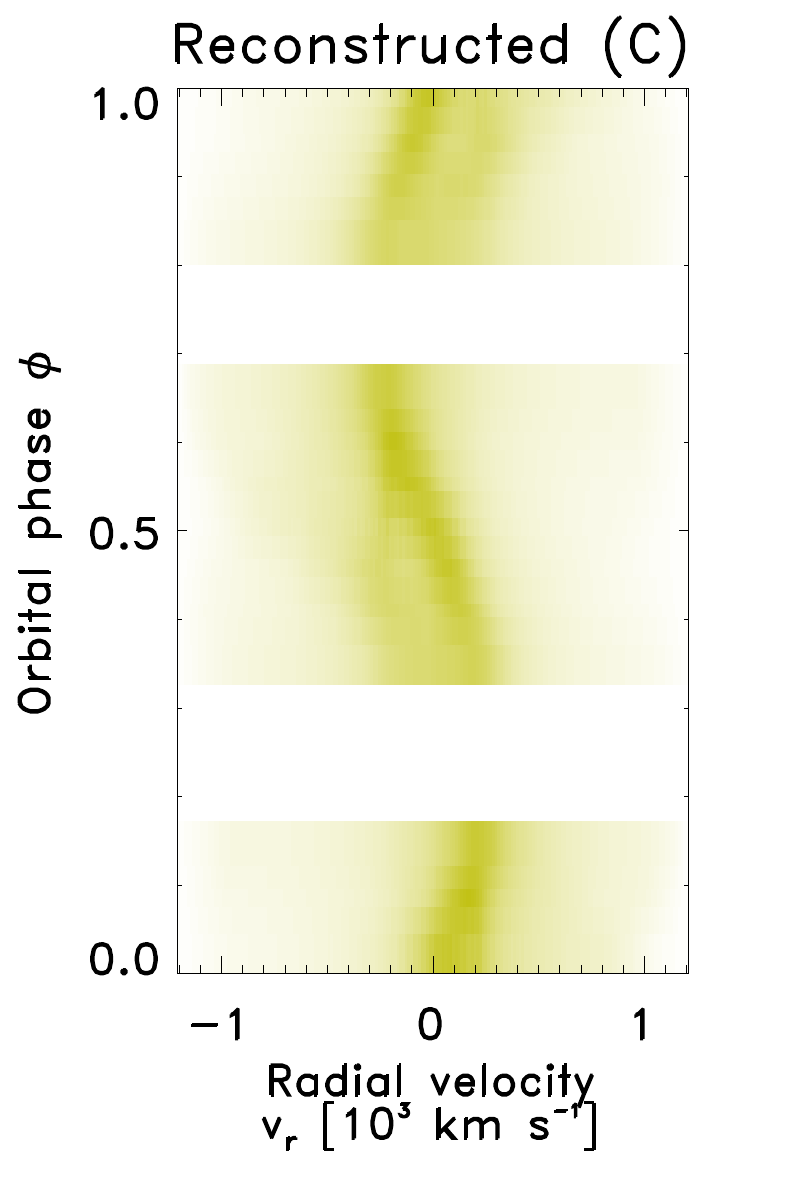} 
 \hspace{-.10cm}
\includegraphics[width=0.15\textwidth, height= 4.5cm]{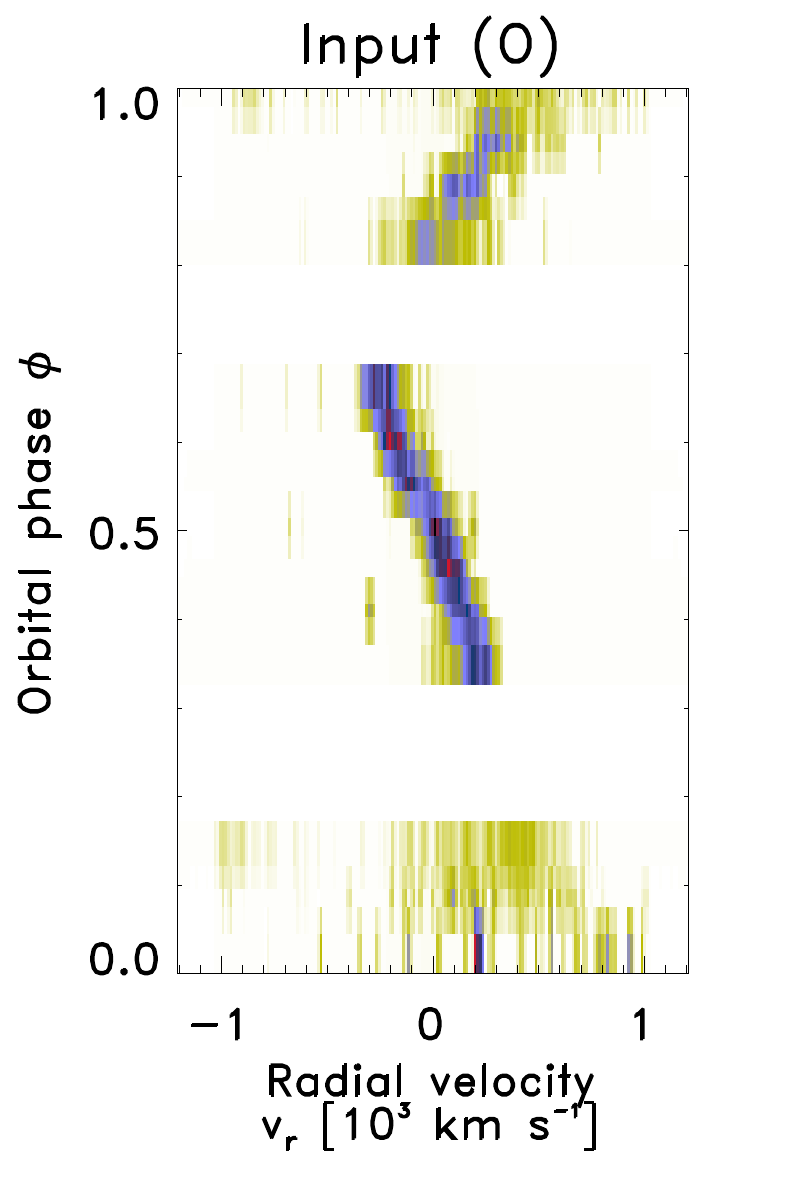} 
 \hspace{-.10cm}
\includegraphics[width=0.15\textwidth, height= 4.5cm]{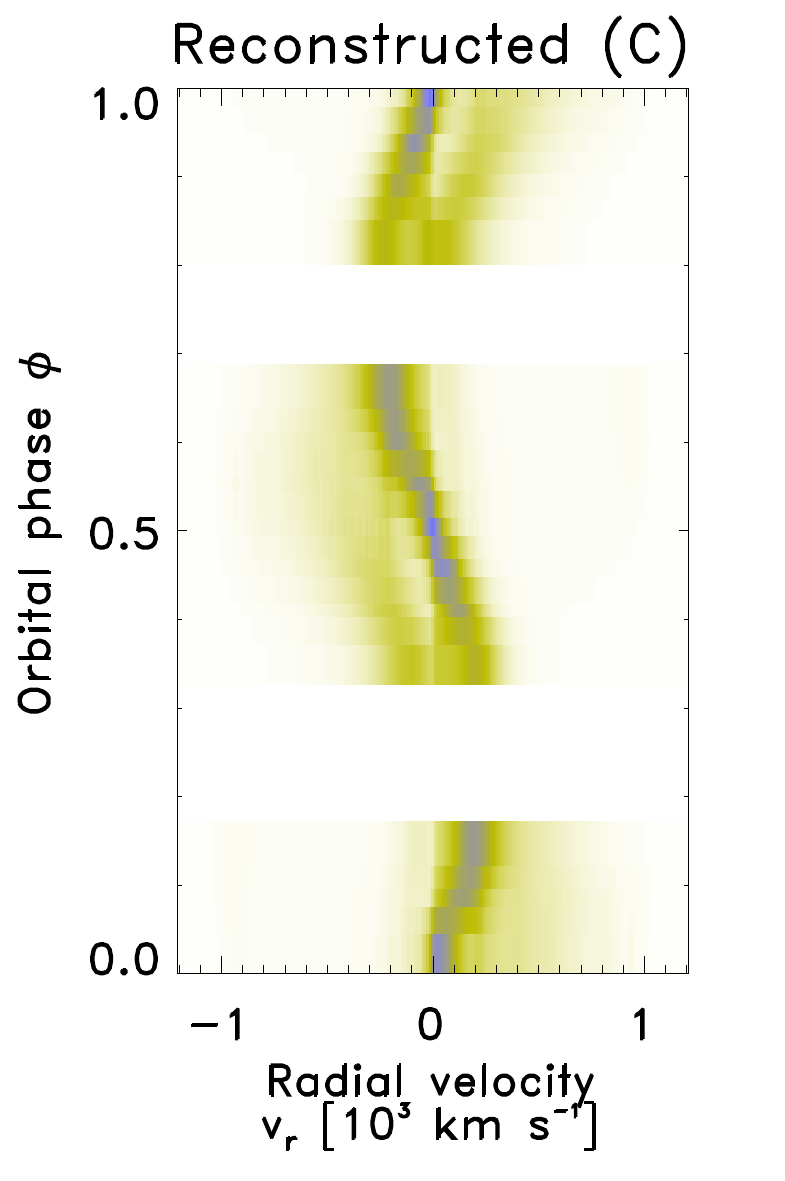}
 \hspace{-0.30cm}
\includegraphics[width=1.0cm, height= 4.50cm]{Spectra_colorbar_20.png} \\ 
\end{array}$
\end{center}
\caption{Same as Fig. \ref{figure:doppsBeta_A} but for CaII 8542 \AA{}.}
\label{figure:doppsCaII8542a}
\end{figure}

The top rows of Figs \ref{figure:doppsBeta_A} to \ref{figure:doppsCaII8542a} show the stand (left panel) and inside-out (right panel) Doppler maps based on the H$\beta$, HeII 4686 \AA{} and CaII 8542\AA{} emission lines, respectively. The bottom rows of the same figures show the corresponding observed (centre) and reconstructed trailed, based on the standard (left) and inside-out (right) projection, of the same lines. 
To aid the interpretation of emission in the tomograms, we have over-plotted a model with WD mass, $M_1$ = 0.71 $M_{\odot}$, mass ratio ($q$ = $\frac{M_2}{M_1}$) of 0.2 and inclination, $i$ = 81$^{\circ}$ \citep{1991MNRAS.253...27B}.  
In the standard projection Doppler maps, binary's centre of mass is marked with a plus (+) sign. The plus sign is also the centre of the map. In the inside-out projection Doppler maps, the centre of mass of the binary is the zero velocity outer circumference of the map. 
The centre of mass of both the primary and secondary are marked with a cross ($\times$) in all our Doppler maps, i.e. both standard and inside-out projections. 
The Roche lobe of the WD is shown with a dashed line whereas that of the secondary is shown with a solid line in both the standard and inside-out Doppler maps. The trajectory of the ballistic stream is marked with a solid line from $L_1$ up to 45$^{\circ}$ in azimuth around the primary. 
The magnetic dipole trajectories are marked with thin dotted blue lines and are calculated at 10$^{\circ}$ intervals in azimuth around the primary. The colour-bars in both figures, to the right of the tomograms, show the scale with which the Doppler maps and trailed spectra were produced. We start by discussing the observed trailed spectra. 

The observed trailed spectra of H$\beta$ and HeII 4686 \AA{} lines show the presence of three distinct emission components. 
The first is a relatively narrow component (red color) which has a low-velocity amplitude and is understood to be associated with the accretion stream. 
The second component is a broad emission line (blue color) which has a mid-velocity amplitude. 
The third component is a relatively broad feature (yellow color) which is visible throughout the orbital phase -- associated with emission produced in different parts of the accretion flow. 
The observed trailed spectra from the CaII 8542 \AA{} does not cover the whole orbital phase but shows evidence of the emission from the narrow and probably broad components from phases 0.3 to 0.7. The narrow component is associated with the emission from the irradiated face of the secondary star. 

The reconstructed trailed spectra of the three lines considered based on both the standard and inside-out projection reproduces the basic structure of the observed trail spectra. Noticeable, the low-velocity component (red) is absent in the reconstructed trailed spectra. 
The observed flux distribution is not reproduced in all the reconstructed trailed spectra. However, the reconstructed trailed spectra based on the inside-out method do show traces of the narrow component.

\subsubsection{Standard and inside-out Doppler tomograms}

The standard Doppler tomograms based on H$\beta$ and HeII 4686 \AA{} are dominated by emission from the threading region, ballistic and magnetic confined stream and the bulk of the emission is centred at velocities of $\sim$500 km/s and $\theta$ of $\sim$160$^{\circ}$. There is little to no evidence of emission from the vicinity of the secondary star in the standard tomograms. 
However, the standard tomogram based on the CaII 8542 \AA{} line shows emission at the position of the secondary star and possibly part of the ballistic stream. 

The inside-out tomograms, based on the H$\beta$ and HeII 4686 \AA{} lines, reveal the presence of emission from two main regions, namely: the threading region, and the ballistic and magnetically confined streams. There is low level emission from the secondary star and this is seen as a diffuse feature covering the Roche lobe of the secondary in both tomograms. 
The ballistic stream (indicated by a solid black line from $L_1$) starts very faint from the secondary star and brightens up on or before reaching the threading region. 
The threading region dominates the emission in both the inside-out Doppler tomograms of H$\beta$ and HeII 4686 \AA{} lines. 
At the threading region, the accretion stream slows down and is deflected to move perpendicular to the motion of the binary along the magnetic field lines. This motion is indicated by the red dotted lines in the inside-out Doppler tomograms. 
Also visible are the various parts of the magnetic confined stream in the third quadrant as the material is forced to move along the magnetic field lines, resulting in the formation of an accretion curtain, before been funneled onto the magnetic pole of the WD. 
The funneling of materials onto the WD is clear in the inside-out Doppler map and more pronounced in the tomogram of H$\beta$ line.


The inside-out tomogram based on the CaII 8542 \AA{} emission line shows a map dominated by emission from the irradiated face of the secondary star as well as part of the ballistic stream. 
We can not say much about the secondary star since the spectra obtained by SALT did not cover a complete orbital cycle and also the fact that the two observations in the red were taken a year apart. However, the CaII line can be used as a tracer for the secondary star.  

\subsubsection{Standard and inside-out modulation amplitude maps}

\begin{figure}
\begin{center}$
\begin{array}{ccc}
\includegraphics[height= 4cm, width=4cm]{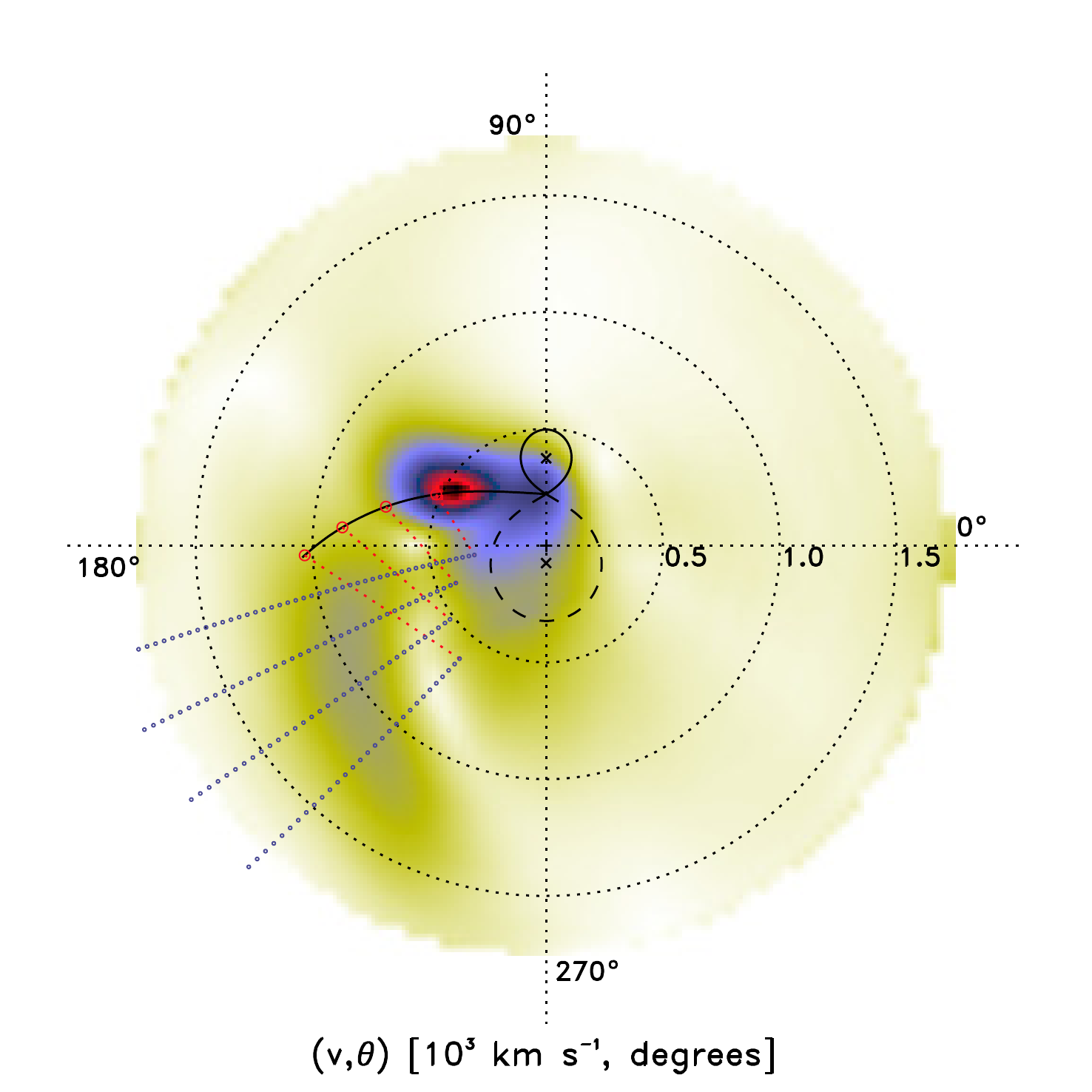} 
 \hspace{-0.25cm}
\includegraphics[height= 4cm, width=4cm]{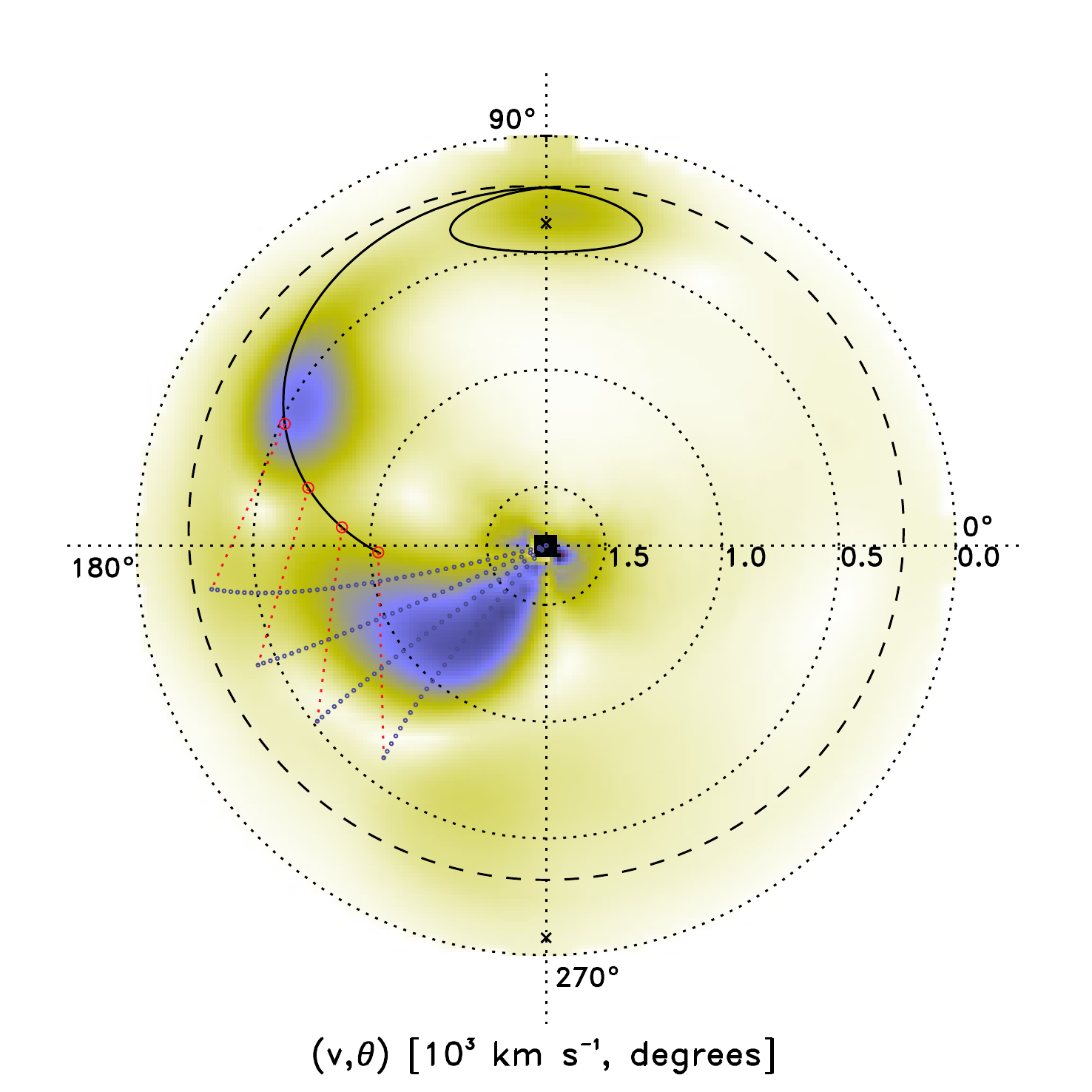}
 \hspace{-0.5cm}
\includegraphics[height= 4cm, width=1.0cm]{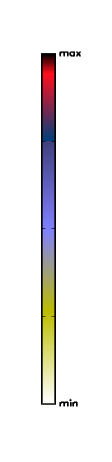} \\ 
\end{array}$
\end{center}
\vspace{-0.35cm}
\begin{center}$
\begin{array}{ccc}
\includegraphics[height= 4cm, width=4cm]{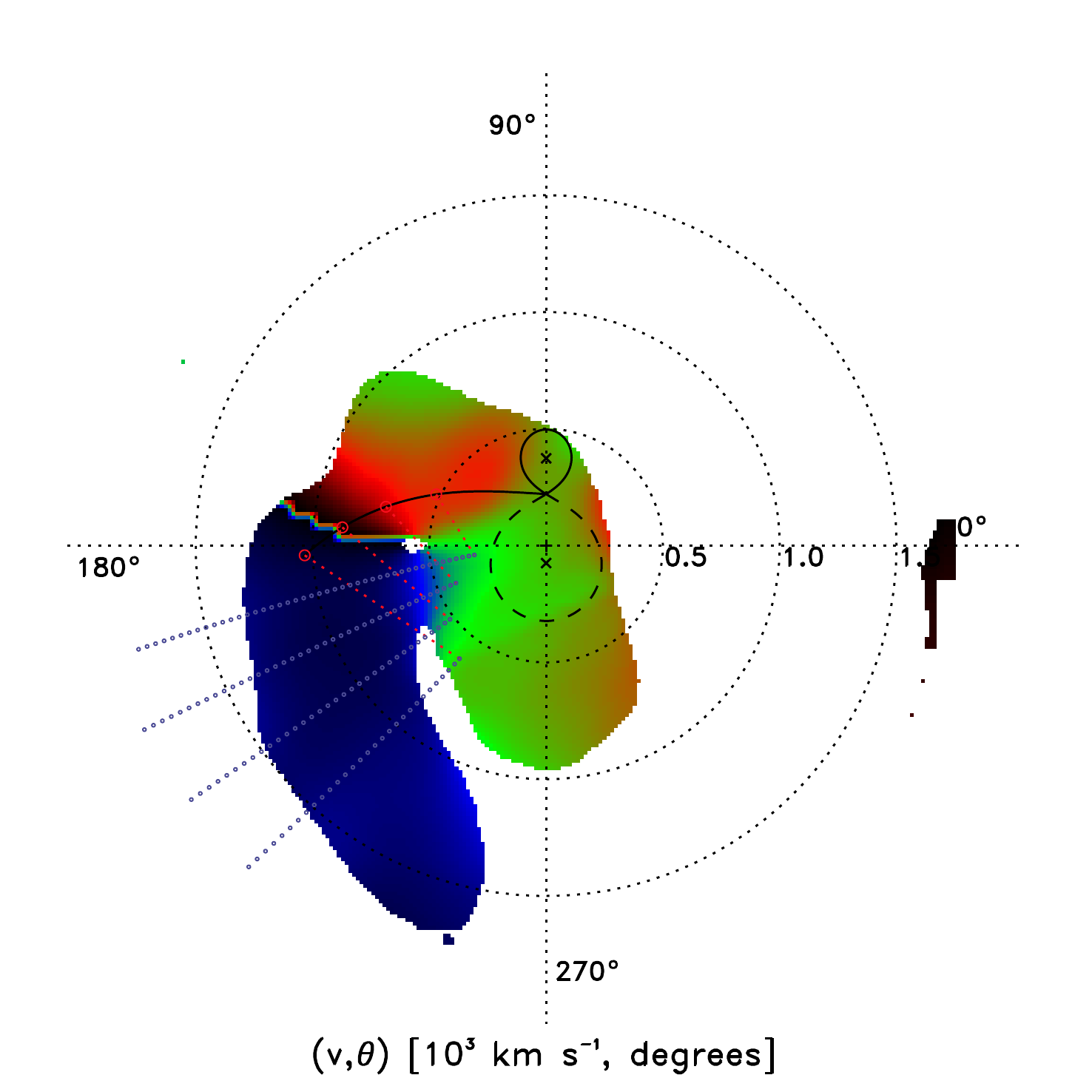} 
 \hspace{-0.250cm}
\includegraphics[height= 4cm, width=4cm]{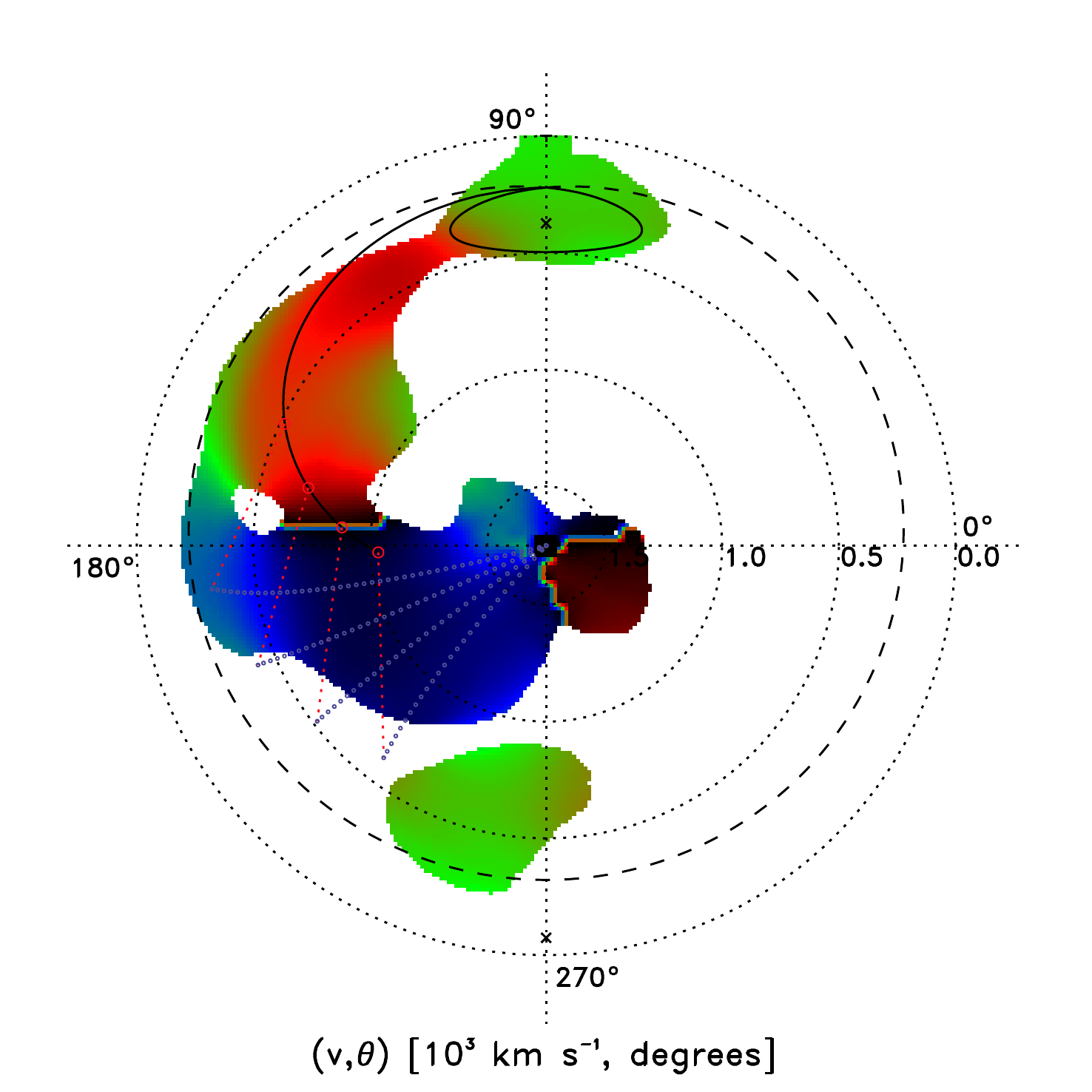}  
\hspace{-0.50cm}
\includegraphics[height= 4cm, width=1.0cm]{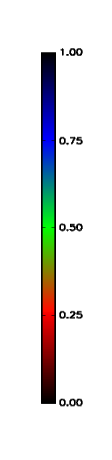}  \\
\end{array}$
\end{center}
\vspace{-0.15cm}
\begin{center}$
\begin{array}{cccc}
\includegraphics[width=0.15\textwidth, height= 4.5cm]{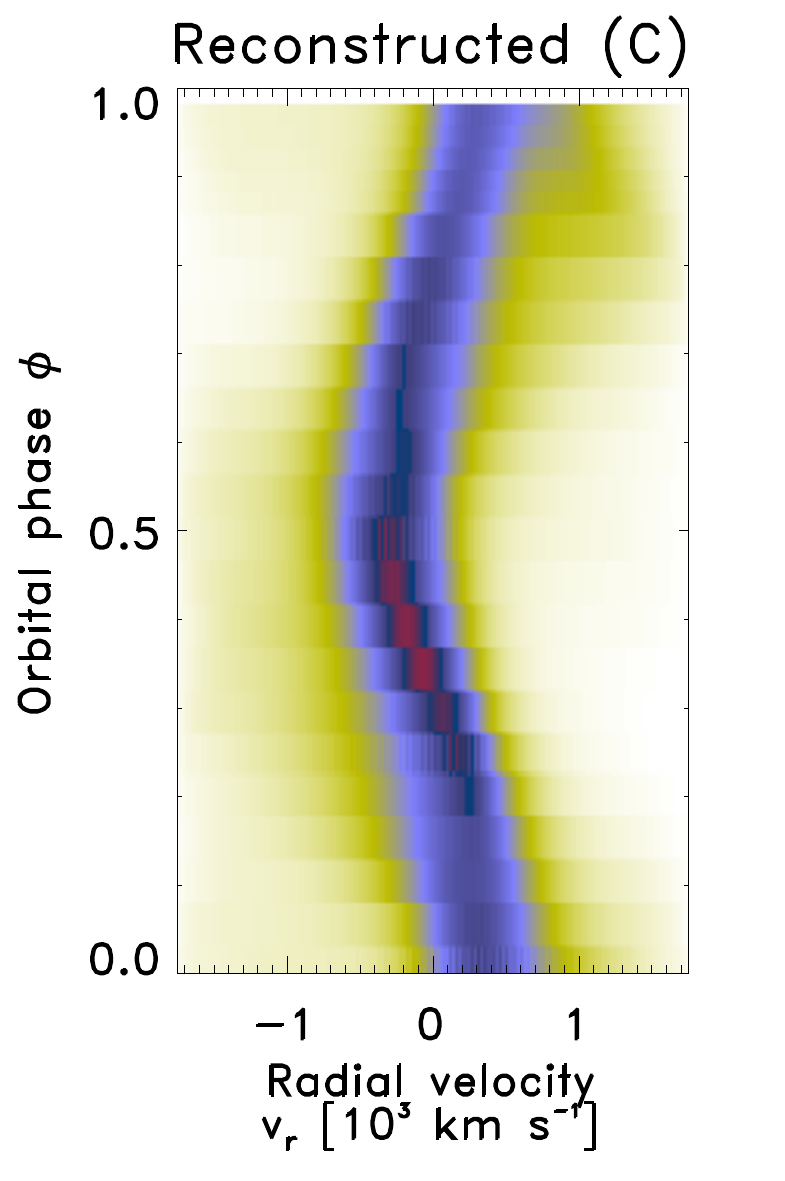} 
 \hspace{-.10cm}
\includegraphics[width=0.15\textwidth, height= 4.5cm]{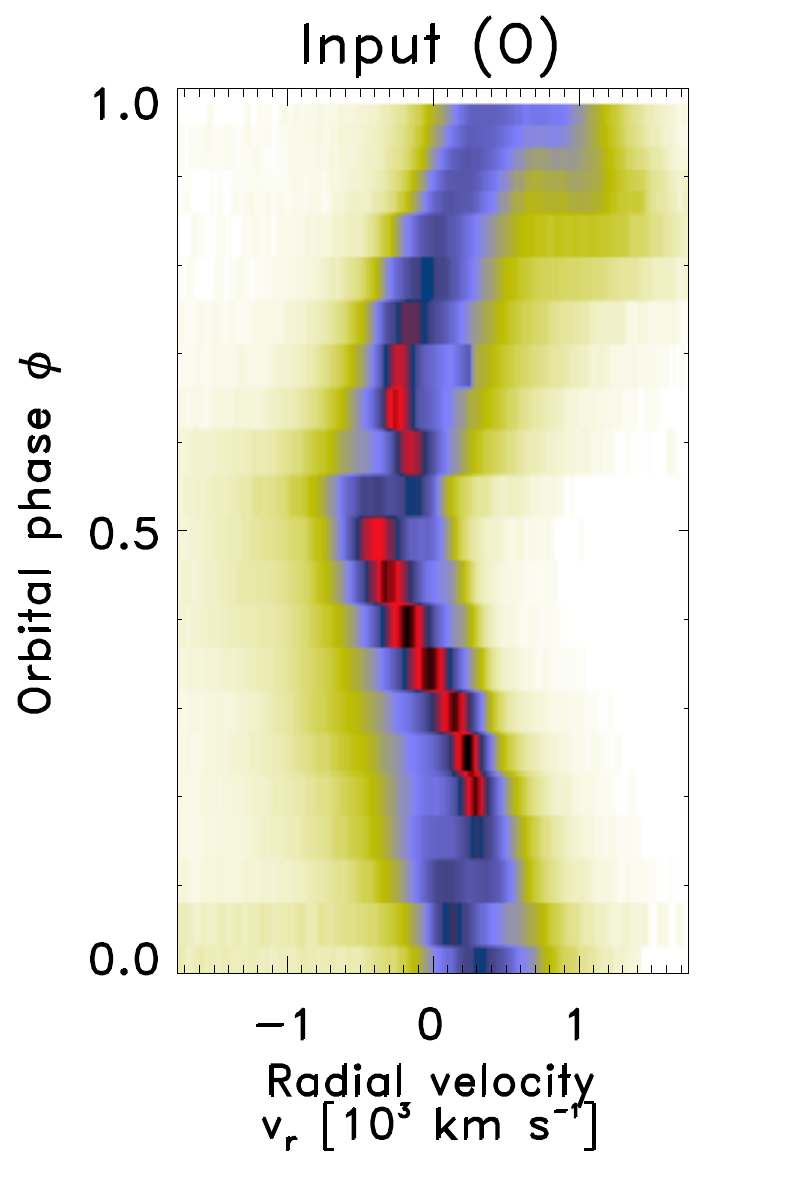} 
 \hspace{-.10cm}
\includegraphics[width=0.15\textwidth, height= 4.5cm]{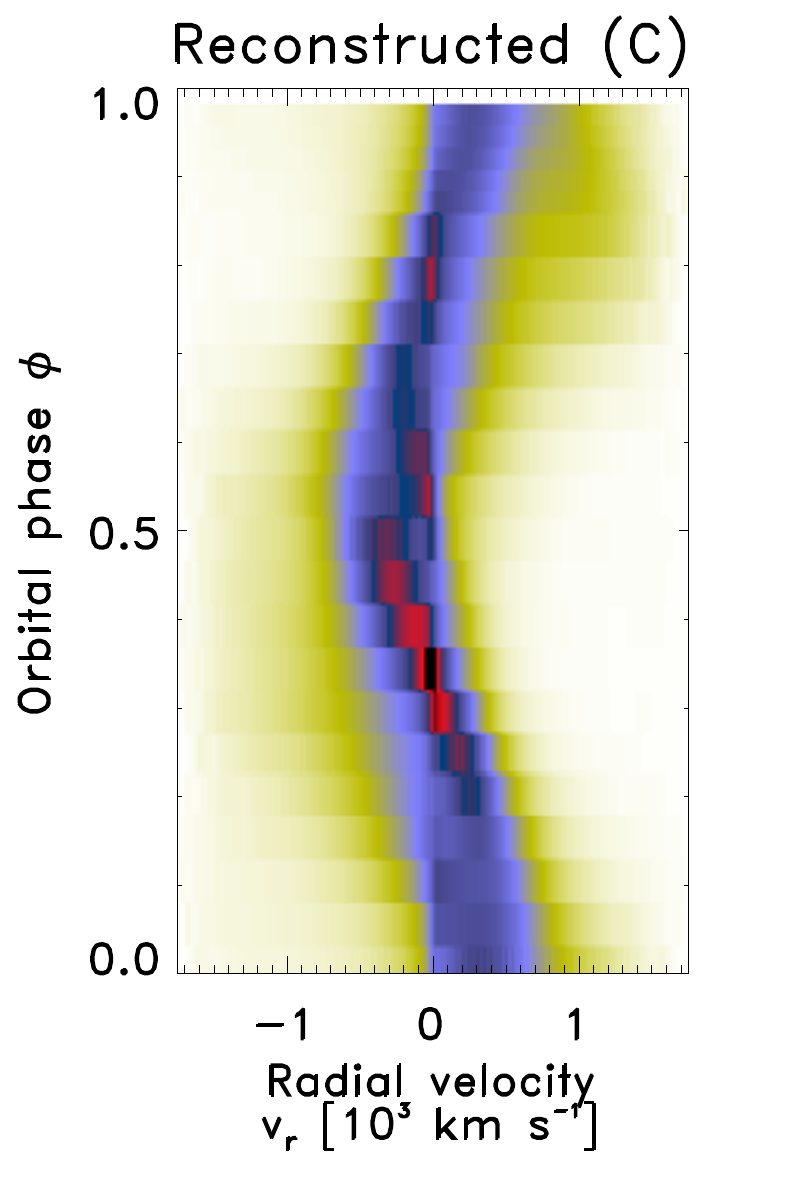} 
 \hspace{-0.30cm}
\includegraphics[width=1.0cm, height= 4.5cm]{Spectra_colorbar_20.png} \\ 
\end{array}$
\end{center}
\caption{Standard and inside-out Doppler maps and trailed observed and reconstructed spectra based on the the $H\beta$ emission line. Top row: the standard and inside-out modulation amplitude flux Doppler maps. Middle row: the standard and inside-out phase of maximum flux Doppler maps. Bottom row: the input trailed spectra (centre) with the summed reconstructed trailed spectra for the ten consecutive half-phases for standard (left) and inside-out (right), respectively.}
\label{figure:doppsBeta_B}
\end{figure}

\begin{figure}
\begin{center}$
\begin{array}{ccc}
\includegraphics[height= 4.cm, width=4.cm]{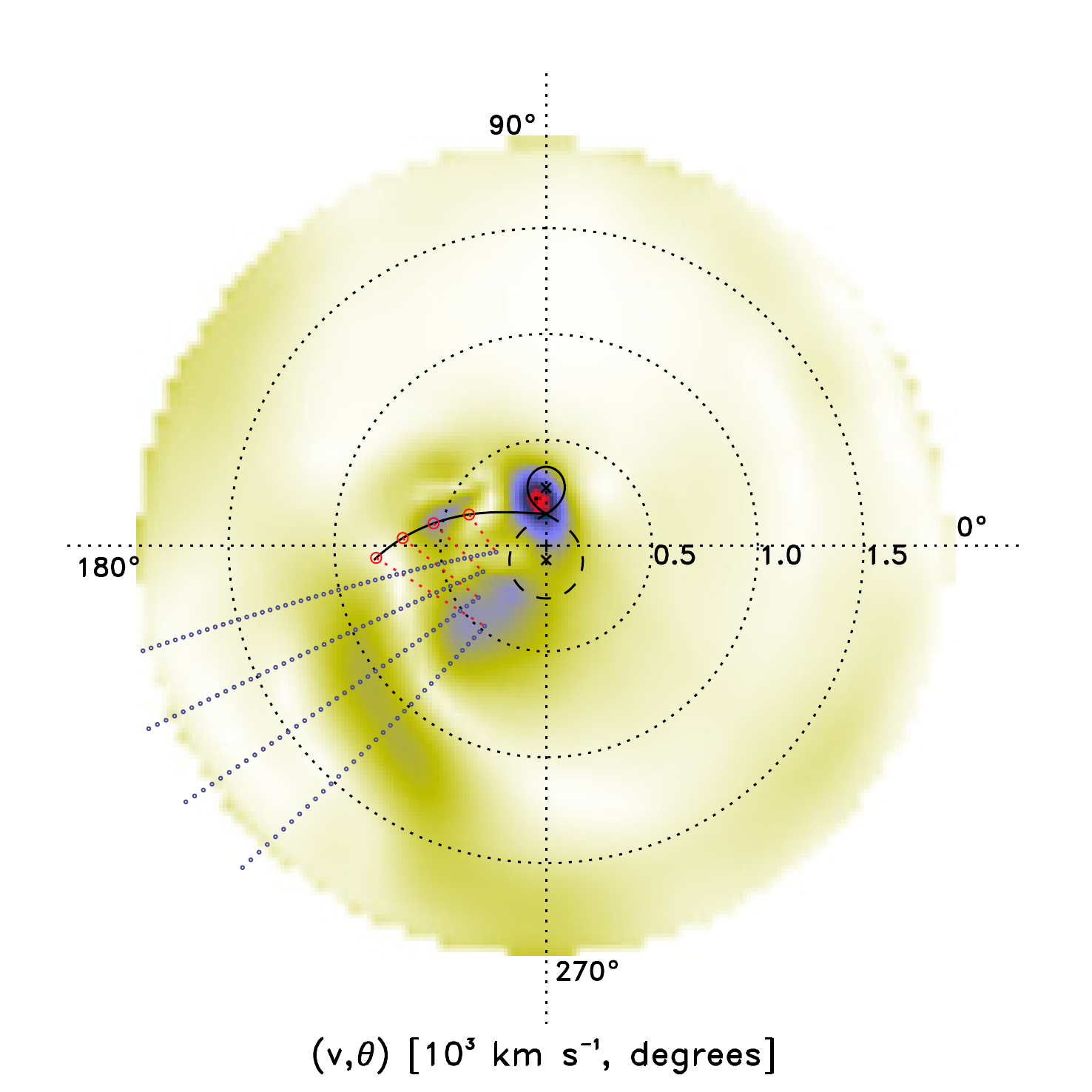} 
 \hspace{-0.25cm}
\includegraphics[height= 4.cm, width=4.cm]{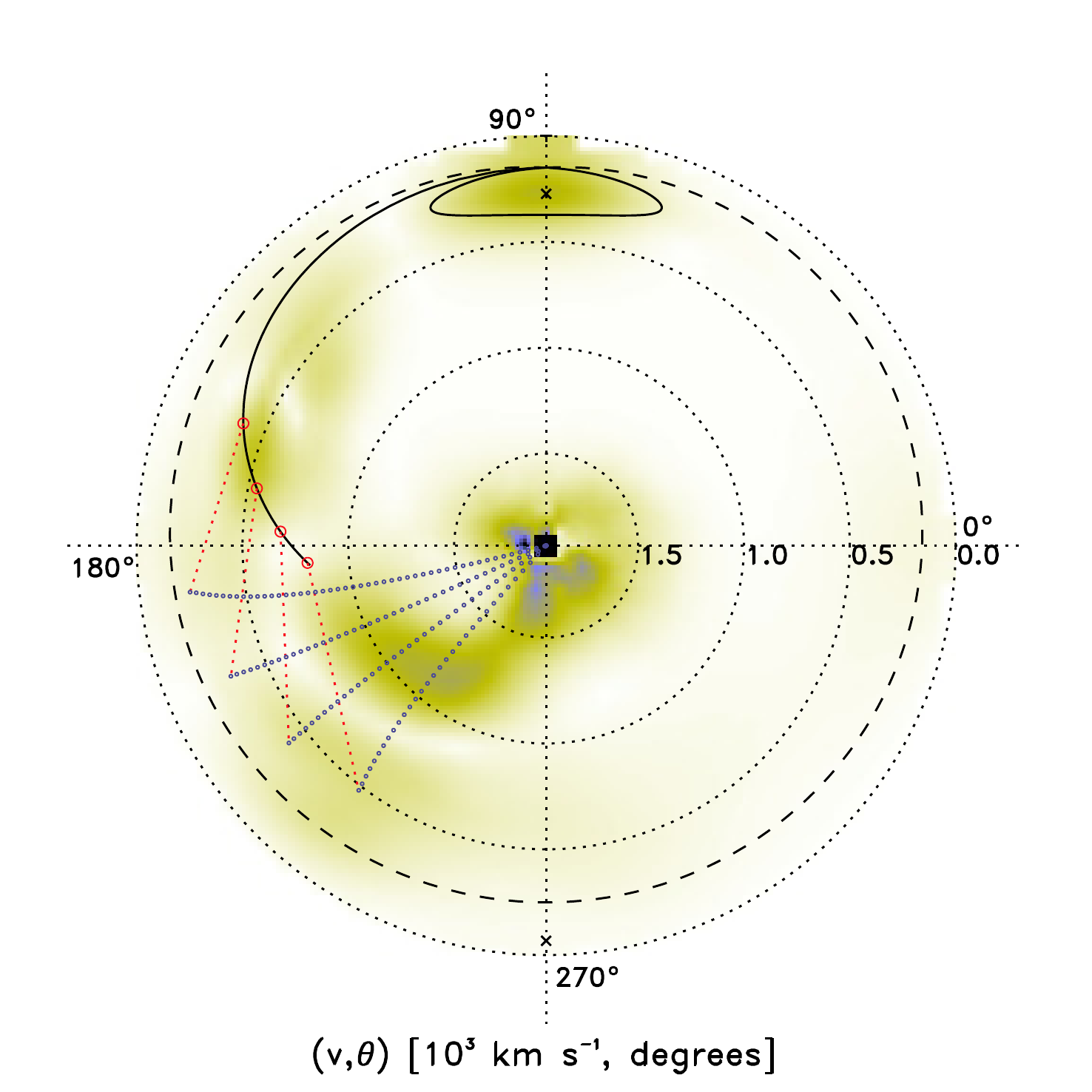}
 \hspace{-.50cm}
\includegraphics[height= 4.cm, width=1.0cm]{Amplitude_colorbar_20.png} \\ 
\end{array}$
\end{center}
\vspace{-0.35cm}
\begin{center}$
\begin{array}{ccc}
\includegraphics[height= 4.cm, width=4.cm]{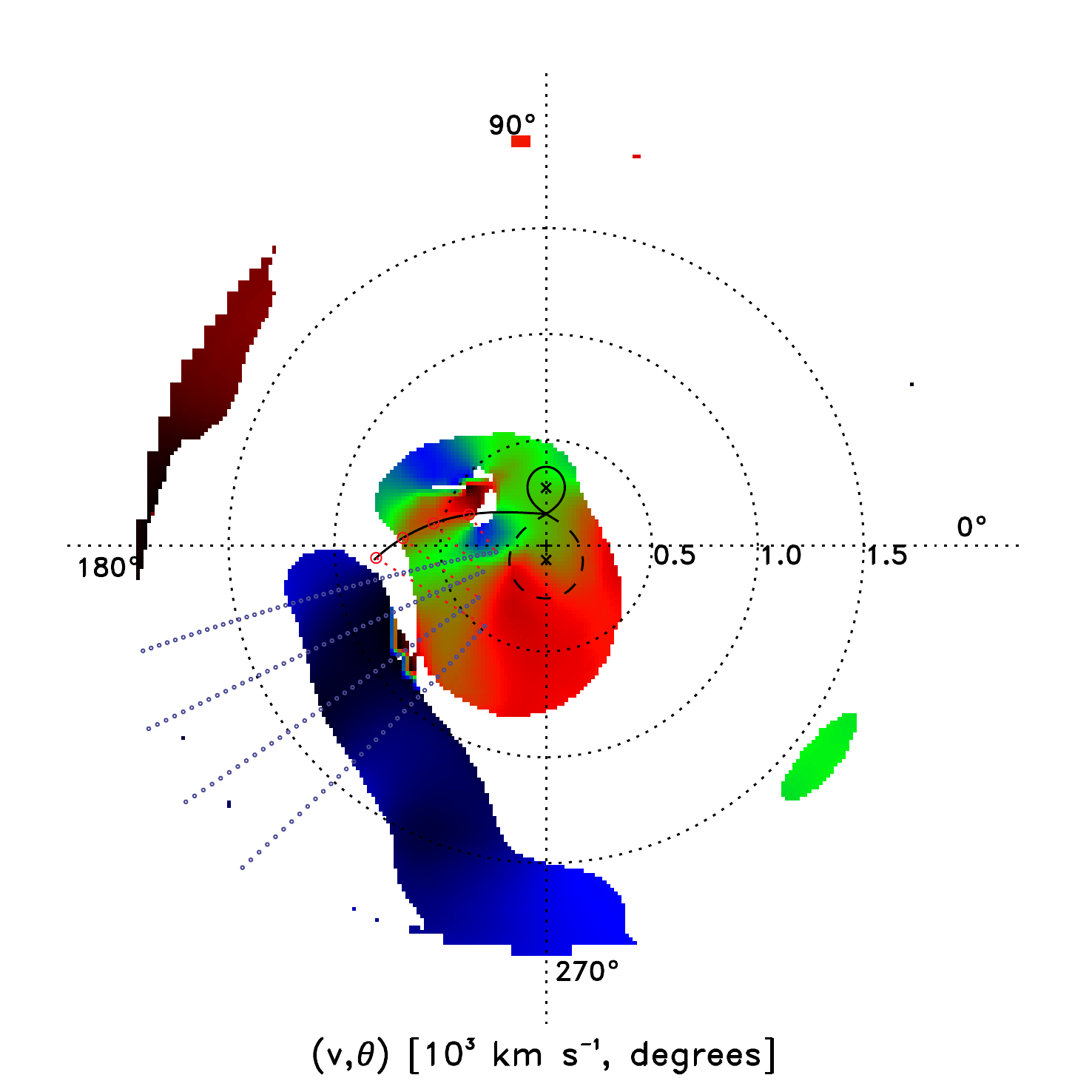} 
 \hspace{-0.250cm}
\includegraphics[height= 4.cm, width=4.cm]{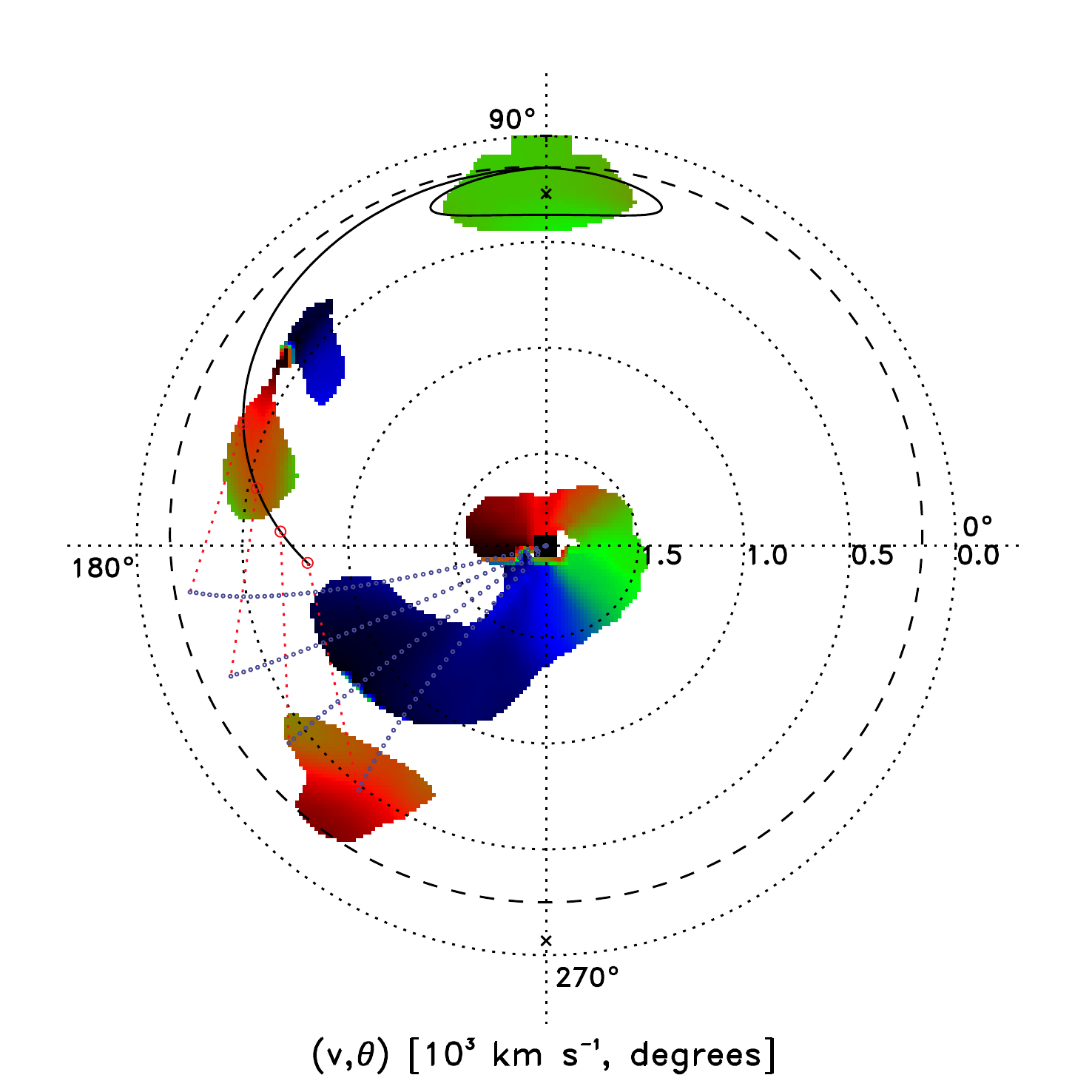}  
\hspace{-0.50cm}
\includegraphics[height= 4.cm, width=1.0cm]{Phase_colorbar_20.png}  \\
\end{array}$
\end{center}
\vspace{-0.15cm}
\begin{center}$
\begin{array}{ccc}
\includegraphics[width=0.15\textwidth, height= 4.5cm]{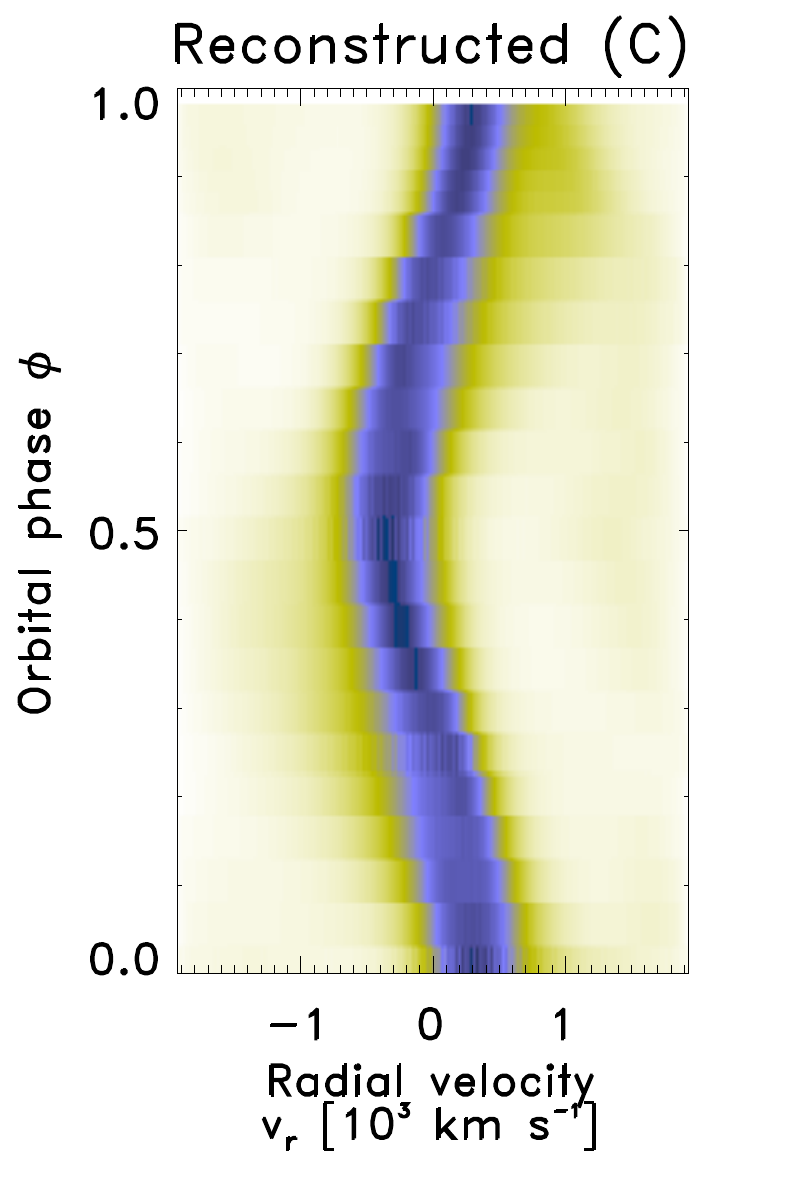} 
 \hspace{-.10cm}
\includegraphics[width=0.15\textwidth, height= 4.5cm]{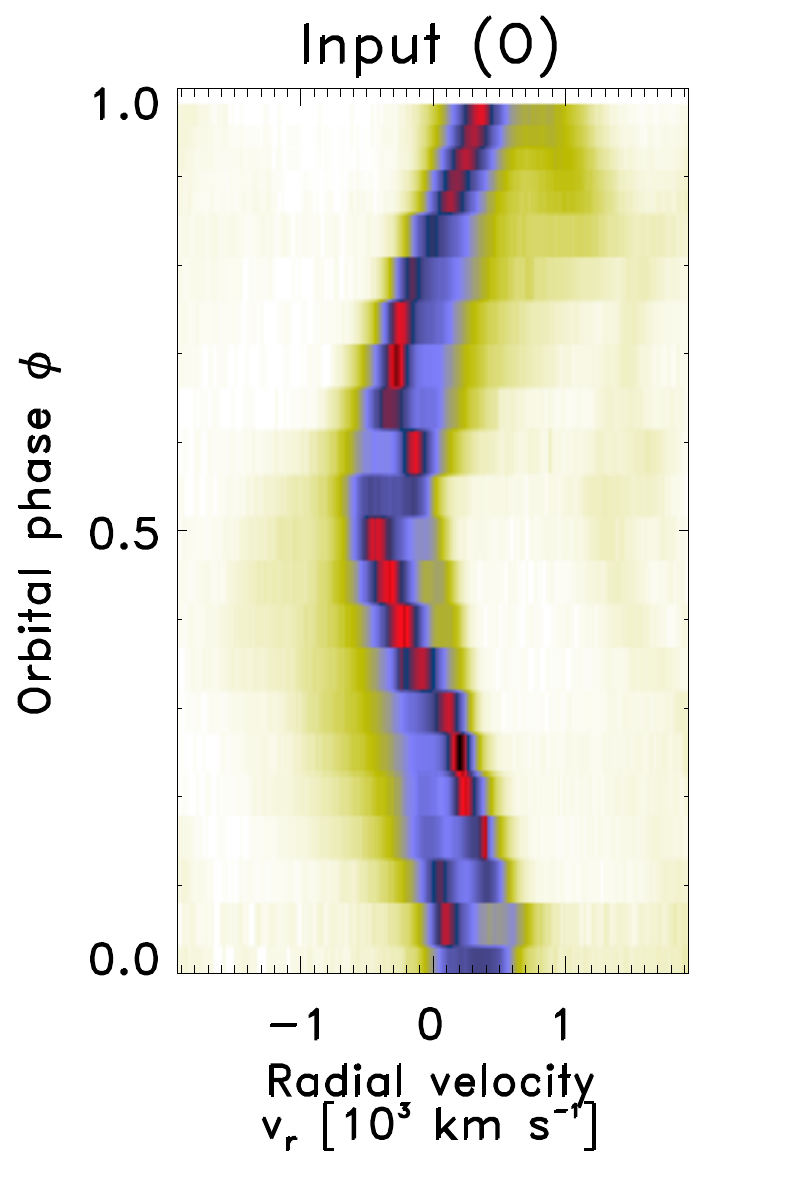} 
 \hspace{-.10cm}
\includegraphics[width=0.15\textwidth, height= 4.5cm]{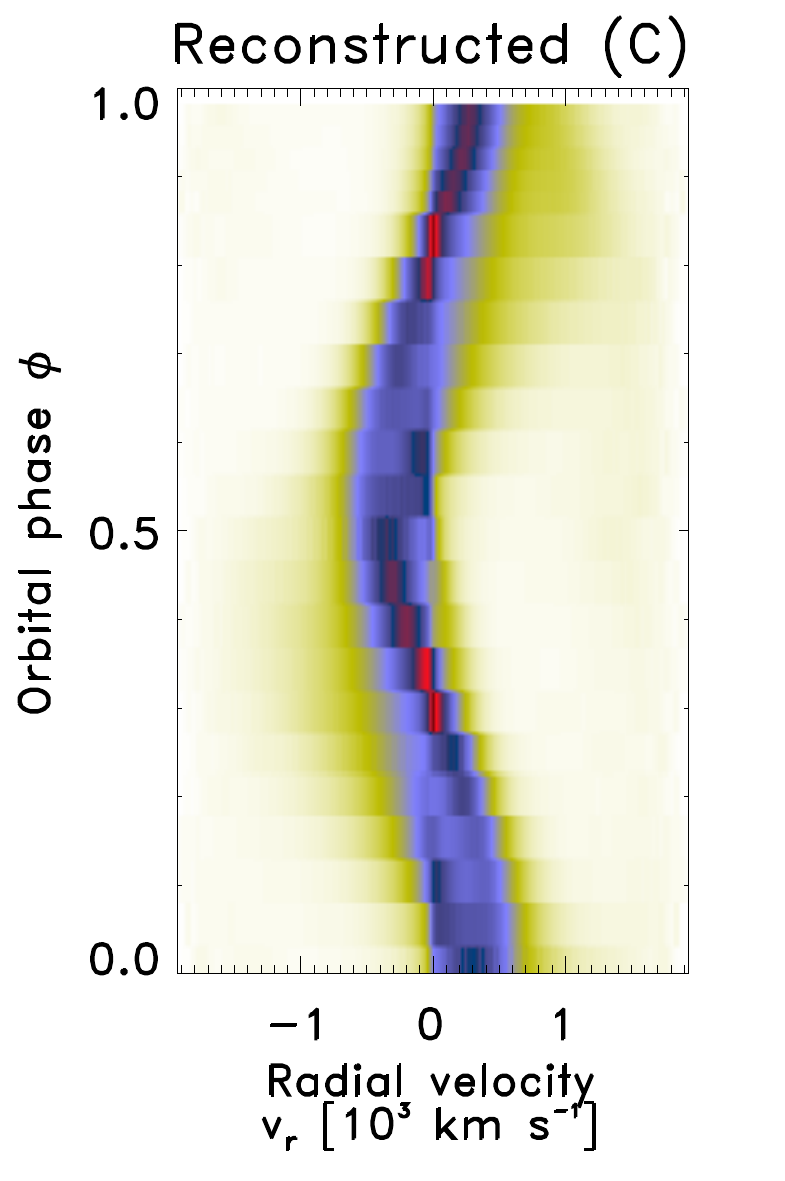} 
 \hspace{-0.30cm}
\includegraphics[width=1.0cm, height= 4.5cm]{Spectra_colorbar_20.png} \\ 
\end{array}$
\end{center}
\caption{Same as Fig. \ref{figure:doppsBeta_B} but for HeII 4686 \AA{}.}
\label{figure:doppsHeII_B}
\end{figure}

We also presented Doppler maps based on the flux modulation mapping technique described in \cite{2016A&A...595A..47K} which exploits the principles introduced by \cite{2003MNRAS.344..448S} and \cite{2004MNRAS.348..316P}. 
The modulation mapping technique produces Doppler maps that represent the average, amplitude and phase of the modulated emission. This is achieved by extracting any phased modulation in the observed flux from a series of consecutive half-phase tomograms. 
The maps presented here and shown in Figs \ref{figure:doppsBeta_B} and \ref{figure:doppsHeII_B} are based on ten consecutive half-phases (i.e., 0.0--0.5, 0.1--0.6, ..., 0.7--0.2,..., etc) of the H$\beta$ and HeII 4686 \AA{} emission lines. 
Also presented are the observed and reconstructed trailed spectra from this method. We note that the reconstructed trailed spectra based on the flux modulation mapping reproduces the observed trailed spectra better than the standard Doppler mapping techniques. 
This is because in the standard Doppler tomography techniques \citep{1988MNRAS.235..269M} the flux from each point in the frame of rotation of the binary is assumed to be constant. 
However, this is not the case in eclipsing CVs, the flux from the typical emission modulate in time and this information is lost when spectral features are mapped in Doppler tomography. 

The top row of Figs \ref{figure:doppsBeta_B} and \ref{figure:doppsHeII_B} show the standard (left) and inside-out (right) modulated amplitude maps for both the H$\beta$ and HeII 4686 \AA{} emission lines, respectively. 
It is clear from both figures that the ballistic and magnetic confined streams are the most flux modulated components between the two projections used. The secondary star is also shown to modulate for both H$\beta$ and HeII 4686 \AA{} lines, but this is only clear in the inside-out projection and is indicated by the yellow patch overlaid on the Roche lobe of the secondary star. Since UZ For is a high-inclination eclipsing system, we expect the flux from the irradiated side of the secondary star and the ballistic stream to modulate over the orbital phase of the binary due to changing viewing angles.

\subsubsection{Phase of maximum flux maps}

The middle row of Figs \ref{figure:doppsBeta_B} and \ref{figure:doppsHeII_B} show the standard (left) and inside-out (right) phase of maximum flux maps based on the H$\beta$ and HeII 4686 \AA{} emission lines. These maps show at which phase an emission component appears brightest to an observer and here we only display pixels where the corresponding modulation amplitude is at least 10\% of the maximum amplitude and they are color coded according to phase: 0.0 -- black, 0.25 -- red, 0.5 -- green and 0.75 -- blue. The phases here were calculated with respect to the photometric ephemeris of \citet{2019A&A...621A..31K}. 
It is clear from both figures that the secondary star is visible around phase 0.5. This is shown in green in both the standard and inside-out phase of maximum flux maps. This is what we expect since the irradiated face of the secondary is pointing towards the observer at phase 0.5. 
The ballistic stream and possibly the trailing side of the secondary star appears brightest to the observer around phase 0.25, this is shown by the red stream leaving the secondary star to the threading region. This is because at phase 0.25, the observer has the full view of the ballistic stream and the portion of the irradiated side of the secondary star. 
The vicinity of the threading region, where the stream interacts with the magnetic field, is a mixture of red and blue because at phase 0.25 and 0.75 this region is visible to the observer and hence  brightest around these phases. 
The magnetic confined stream (blue) appears brightest around phase 0.75 to the observer and is shown by a large blue patch filling the third quadrant for both the standard and inside-out Doppler maps.

\subsection{Photopolarimetry}\label{sect:polar}

\begin{figure}
    \centering
    \includegraphics[ width = 0.5\textwidth]{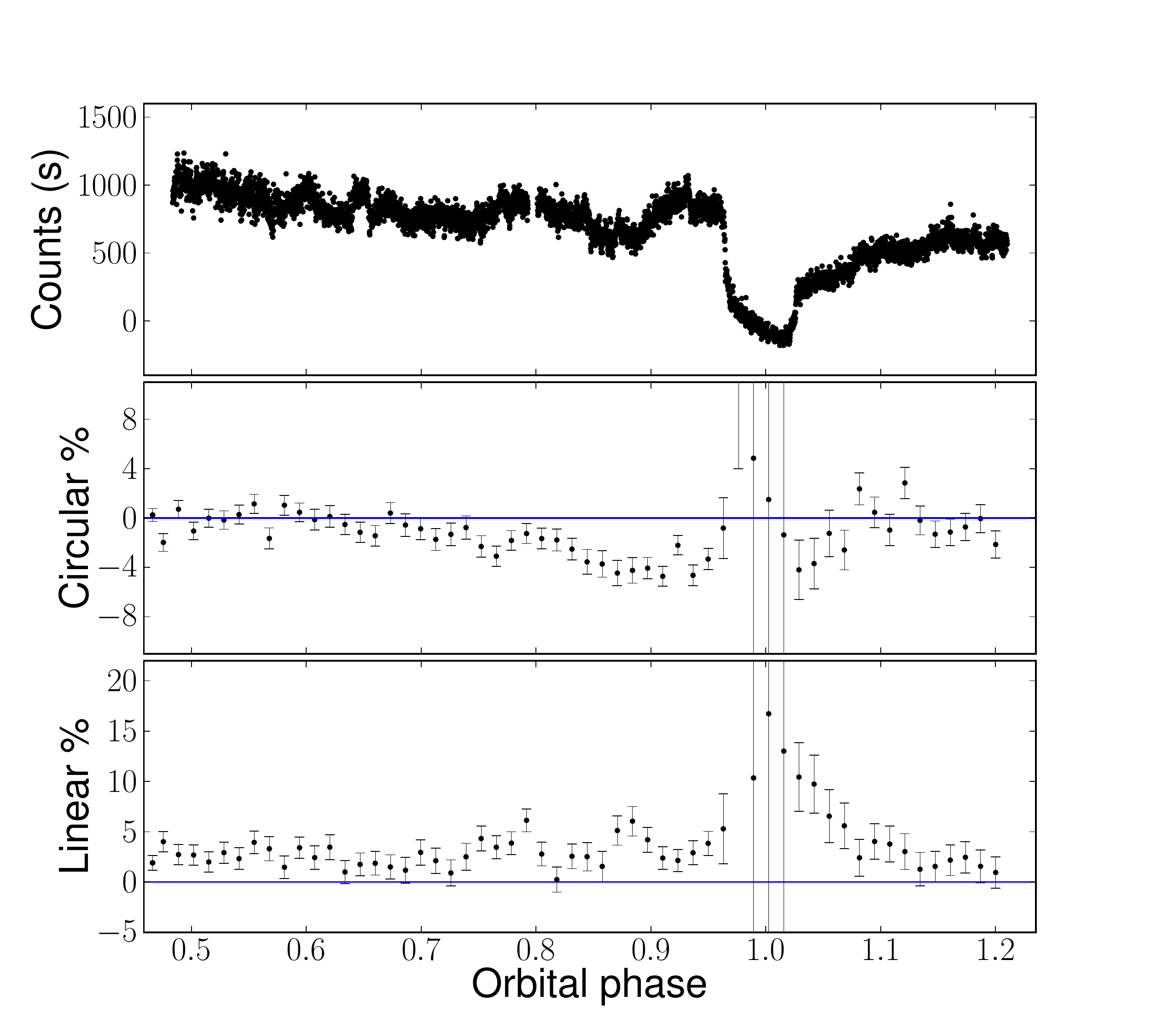}
    \vspace{-0.5 cm}
    \caption{Photopolarimetry from 2018 October 4 made with the HIPPO instrument. Top to bottom panels correspond to photometry, percentage circular and linear polarization.}
    \label{figure:pol_lc}
\end{figure}

Figure \ref{figure:pol_lc} (top panel) shows the phased light curve obtained with the HIPPO instrument. The duration of the light curve is 1.53 hours. The shape of the eclipse is similar to that shown in Fig. \ref{figure:shoc} with clear defined ingress and egress of the main accretion spot. The out-of-eclipse variability is consistent with low amplitude flickering seen in Fig. \ref{figure:shoc}. 
The clear-filtered circular polarimetric observations (Fig. \ref{figure:pol_lc}, middle panel) show variability between 0 and -5\%. The out-of-eclipse circular polarization between phases 0.5 and 0.7 is consistent with zero. Before the eclipse, from phase 0.7--0.95, the polarization increases to -5\%. 
This is because the region emitting cyclotron radiation is visible to the observer around these phases. During the eclipse, the total flux decreases resulting in large error-bars for polarization. After the eclipse, from phases 1.03--1.07, the emission is still negatively polarized with polarization ranging from 0 to -5\%. After phase 1.07, UZ For shows a mixture of polarization which are consistent with zero. 

The bottom panel of Fig. \ref{figure:pol_lc} shows the percentage of linear polarization. The clear-filtered linear polarimetry shows variability between 0 and 10\%. The out-of-eclipse linear polarization, from phase 0.5--0.7, is less than 5\%. Before the eclipse, phases 0.7--0.95, the level polarization increases by a few percent (<10\%) -- consistent with the circular polarization. 
During the eclipse, the total flux decreases resulting in large error-bars for polarization. After the eclipse, there is a pulse of linear polarization reaching about 10\% and decreasing gradually before flattening out between phase 1.1--1.2 and beyond.    

\subsection{Circular spectropolarimetry}\label{sect:spectro}


\begin{figure*}
\centering
 \subfigure[Before the eclipse ($\phi$ = 0.89--0.93)]{
   \includegraphics[width = 0.48\textwidth ]{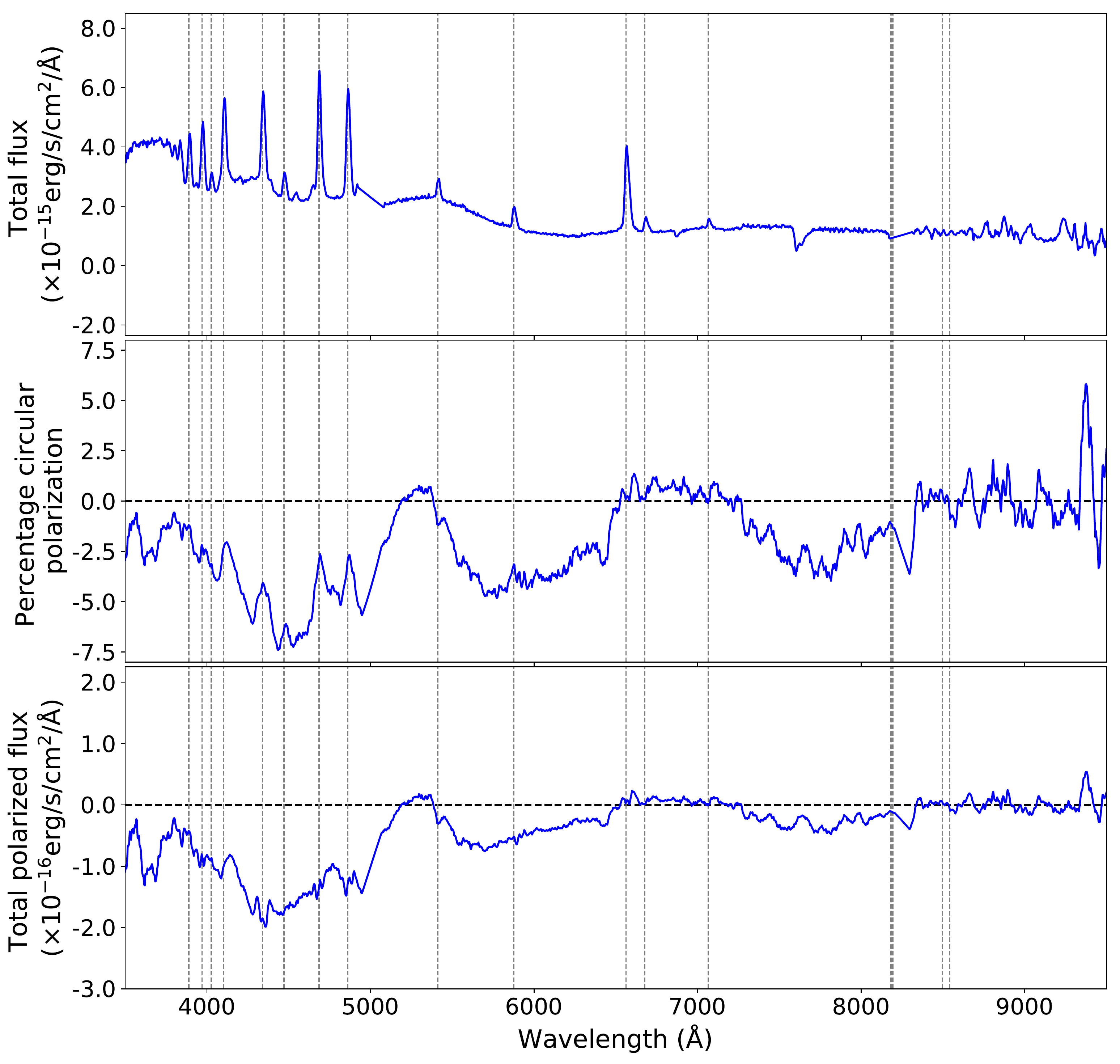}
   \label{fig.1}
   }
 \subfigure[During the eclipse ($\phi$ = 0.97--1.02)]{
   \includegraphics[width = 0.48\textwidth ]{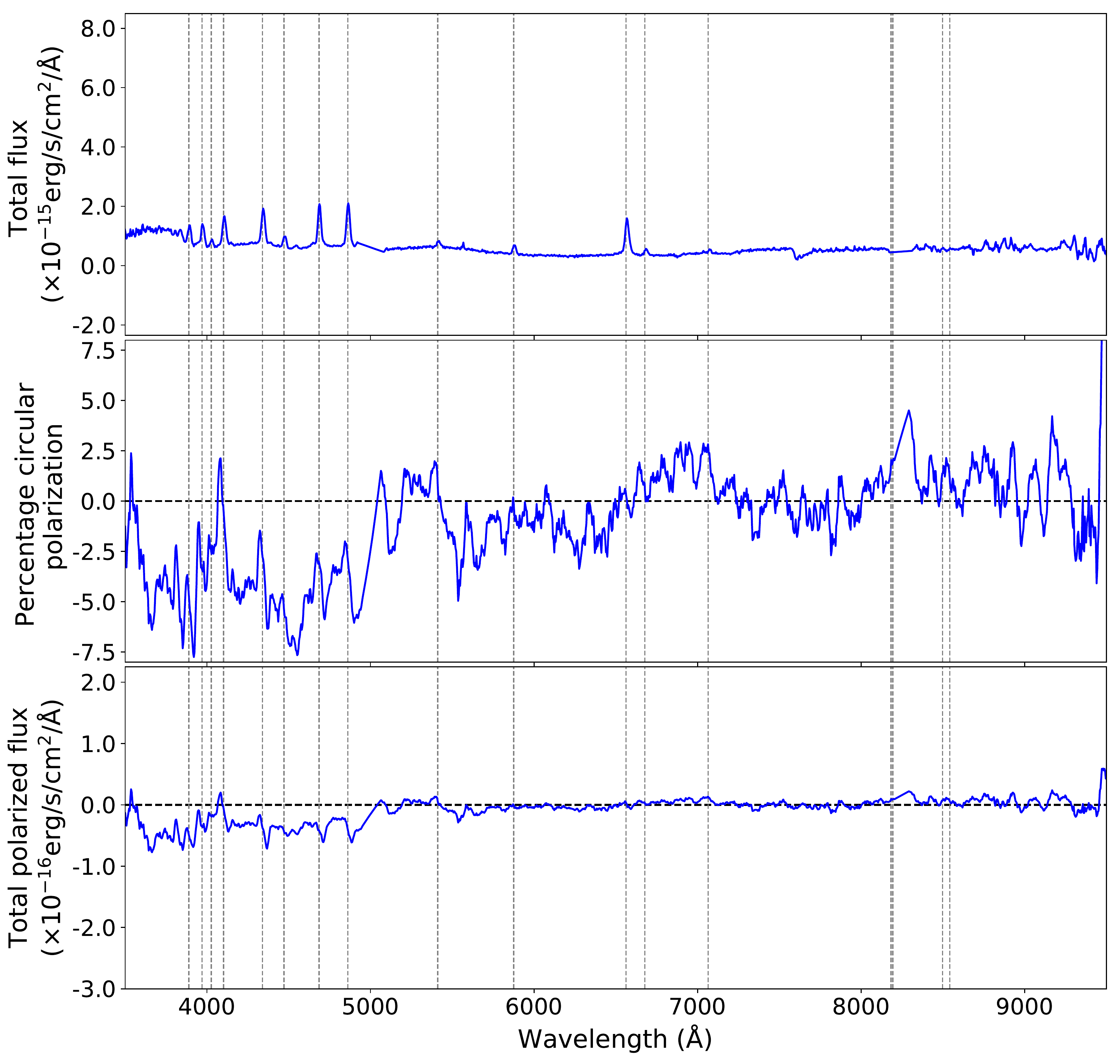}
   \label{fig.2}
   }
 \subfigure[Emerging out of eclipse ($\phi$ = 1.06--1.10)]{
   \includegraphics[width = 0.48\textwidth ]{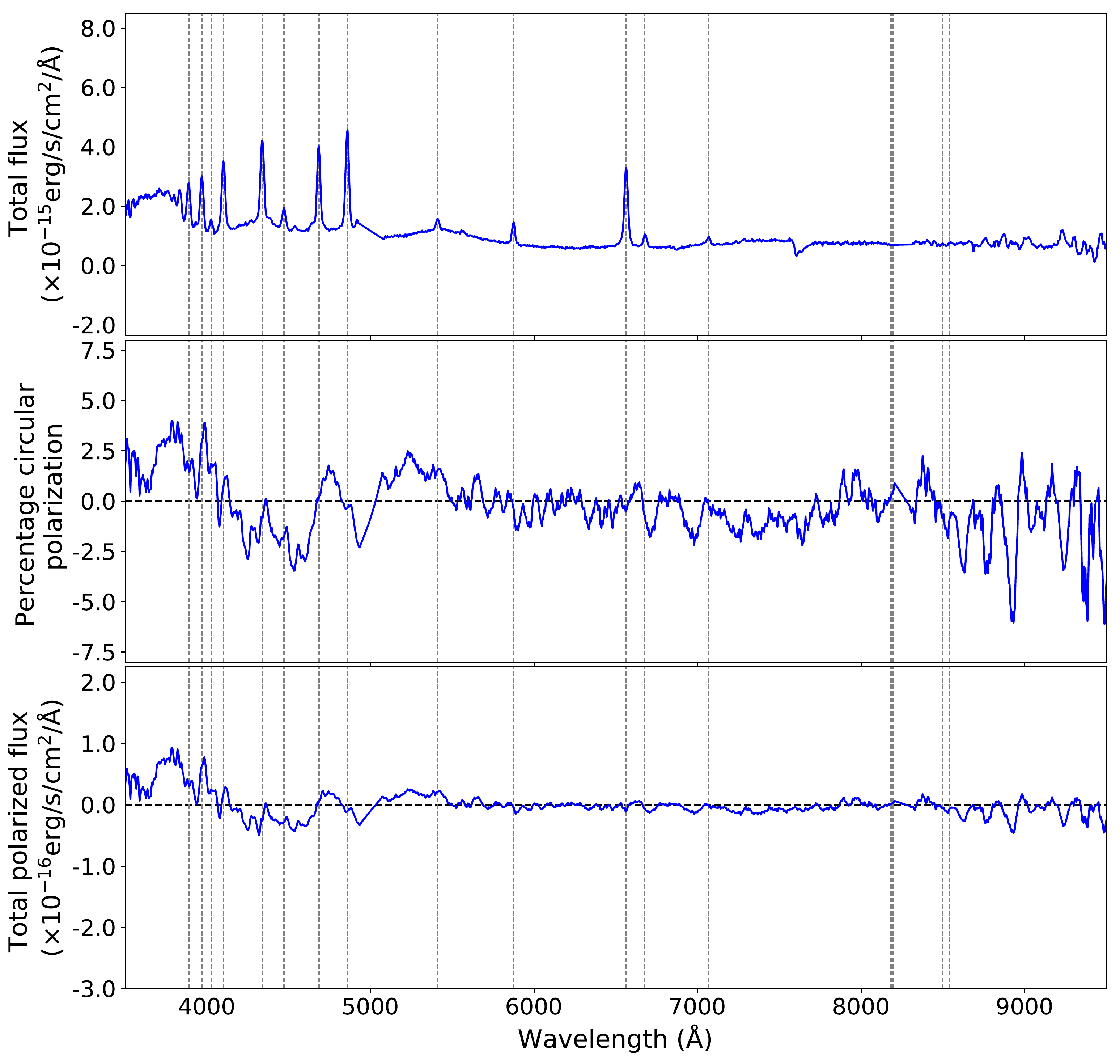}
   \label{fig.3}
   }
 \subfigure[After the eclipse ($\phi$ = 1.14--1.18)]{
   \includegraphics[width = 0.48\textwidth ]{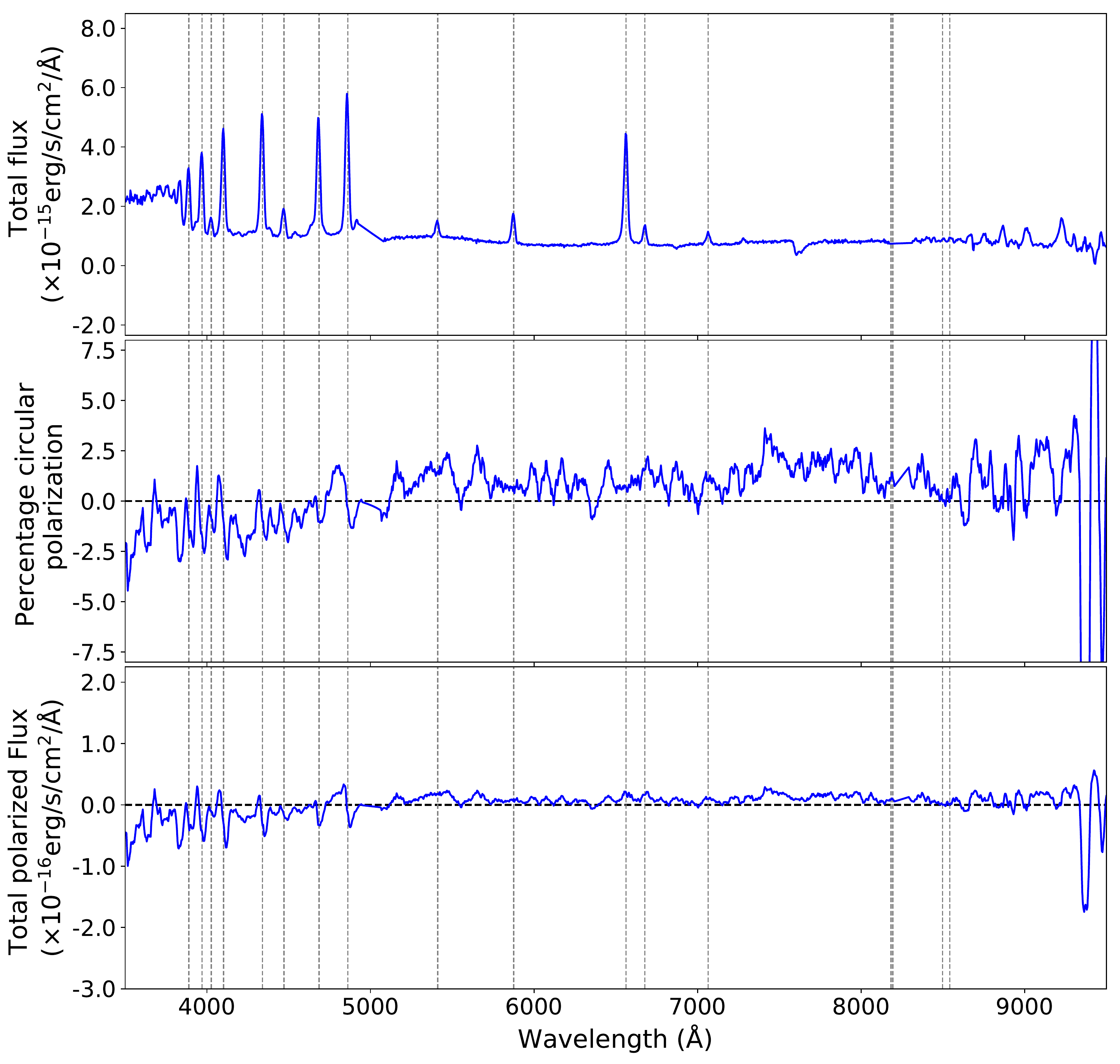}
   \label{fig.4}
   }
\caption{Spectra showing cyclotron emission lines in total flux (top panel), circular polarization (middle panel) and the total polarized flux (bottom panel) of UZ For.}
\label{figure:circ1}
\end{figure*}


\begin{figure}
\centering
\includegraphics[ width = 0.48\textwidth ]{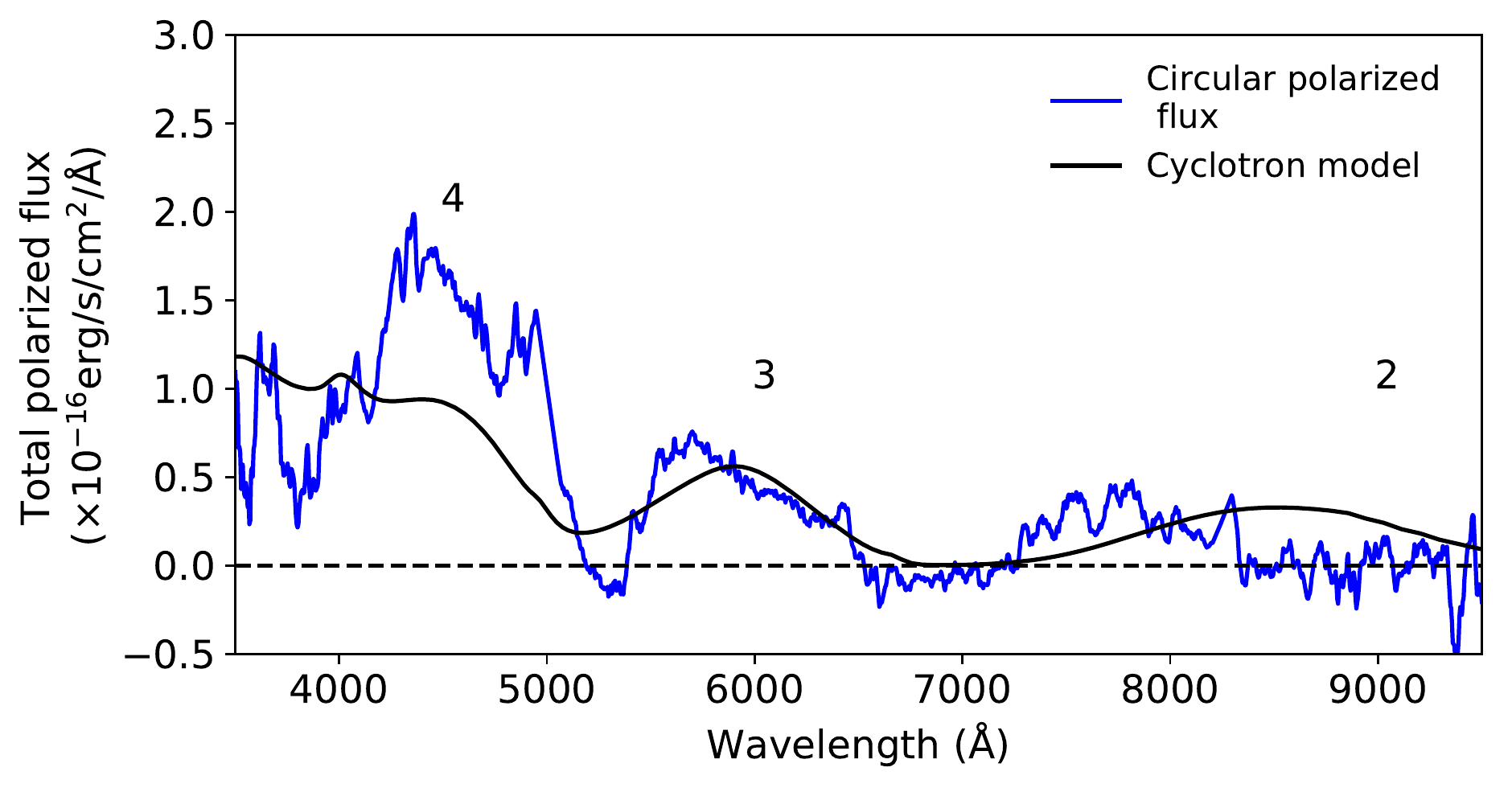}
\caption{Total circularly polarized flux (blue) of UZ For and overlaid is the pure cyclotron model (black) with the magnetic field of 57 MG. The numbers 4, 3 and 2 marks the theoretical positions of the three harmonic features. }
\label{fig:circ_model}
\end{figure}

Figures \ref{fig.1} to \ref{fig.4} show the time-sequence of circular spectropolarimetry obtained before the eclipse, during the eclipse, emerging out of the eclipse and after the eclipse. 
Each panel of the figure (from top to bottom), shows the total flux spectra, the percentage of circular polarization and the total circularly polarized flux are shown. 
The total flux spectra show a continuum which rises in the blue and is dominated by broad emission features covering the entire waveband. As is expected, the total flux is higher before the eclipse and lower during the eclipse when the WD is eclipsed. The total flux again increases when emerging out of eclipse and remain high after the eclipse. 
The spectral features shown are similar to those presented in Sect. \ref{sect:spec_an} and the strength of the emission lines vary throughout the observation. The total flux spectra possibly show a broad hump around 5500 \AA{}.      

\subsubsection{Percentage circularly polarized spectra}

The middle panels of Figs \ref{fig.1} to \ref{fig.4} show the time-sequence of the percentage of circularly polarized spectra of UZ For. They show strong negative circular polarization (up to -8\%) in the blue and decreasing gradually towards the red. 
The grey dashed vertical lines mark the location of the emission lines (as seen in the top panels) and it is clear that there are excursions towards 0\% at their locations since emission lines are not polarized. 
The circular polarization spectra show the presence of three negative polarized humps, centred at $\sim$4500, 6000 and 7800 \AA{}, that are interpreted as cyclotron harmonics due to cyclotron emission from a hot plasma. 
The harmonics are more visible in Fig. \ref{fig.1} (middle panel) before the eclipse. 
During the eclipse, Fig. \ref{fig.2} middle panel, the strength of the harmonics are significantly reduced, especially at longer wavelengths. The reason we see the harmonics during the eclipse is due to that the first exposure started before the ingress time and the second exposure was taken during mid-eclipse. 
When the system is emerging out of the eclipse, Fig. \ref{fig.3}, there is some polarization in the blue part of the spectra. 
After the eclipse (Fig. \ref{fig.4} middle panel) over most of the observed wavelength, the circularly polarized spectra are consistent with 0\% circular polarization. But between 7000 and 9000 \AA{}, there is a marginal detection of positive polarization. 

The circularly polarized spectra of UZ For is understood in terms of Fig. \ref{figure:pol_lc} in that from phases 0.7--0.95 the accretion spot emitting cyclotron radiation is visible to the observer around this phases. The spot is eclipsed between phases 0.95--1.03 and therefore no polarization is observed. 
According to Fig. \ref{figure:pol_lc}, there should be negative circular polarization when the WD emerges from the primary eclipse, but it is not clear whether this is seen in Figs \ref{fig.3} and \ref{fig.4}. However, Fig \ref{fig.3} shows evidence of a negative hump in the blue at 4500 \AA{}. 
Furthermore, there is a marginal detection of positive polarization in Figs \ref{fig.4} towards the red. 

\subsubsection{Total circularly polarized flux}

We multiplied the total flux spectra by the percentage of circularly polarized spectra to get the total circularly polarized flux. The total polarized flux is of pure cyclotron origin and is free of contamination, e.g., emission from the secondary star. The results are shown in the bottom panels of Figs \ref{fig.1} to \ref{fig.4}. 
As expected, much of the circularly polarized flux is seen in the blue in Fig. \ref{fig.1} (bottom panel) just before the eclipse and where the percentage of polarization reaches $\sim$-8\%. During the eclipse (Fig. \ref{fig.2}, bottom panel), some polarized flux is still seen in the blue end of the spectra. 
This implies that either the accretion spot emitting cyclotron radiation is not completely eclipsed during the primary eclipse or there is second region also emitting cyclotron radiation.  
After the eclipse, Figs \ref{fig.3} and \ref{fig.4}, little or zero polarized flux is seen and the continuum is much flatter implying that the accretion spot emitting cyclotron radiation has moved away from the line of sight of the observer. The marginal positive polarization seen (in Fig. \ref{fig.4}) after the eclipse could be coming from the second pole that might also be emitting cyclotron radiation. 

\subsubsection{Modelling the circularly polarized flux}

The circularly polarized flux at phase 0.91 (Fig. \ref{fig.1}, bottom panel) shows broad features which peaks at approximately 4500 \AA{}, 6000 \AA{} and 7800 \AA{}. These features display the characteristic properties predicted by the theory for cyclotron emission from a hot plasma \citep{1982MNRAS.198..975W}.   
At low temperatures, we know that the positions of the $nth$ harmonic for a given magnetic field ($B$) and viewing angle ($\theta$) is given by the following equation: 
\begin{equation}
\label{equ:harm}
\lambda_n = \frac{10710}{n} \left( \frac{10^8 \rm{G}}{B}\right){\rm{sin}\theta}, 
\centering
\end{equation}
\noindent
where $\lambda_n$ is the wavelength of the peak of the harmonic and $n$ is the harmonic number. 

In order to determine the strength of the magnetic field of UZ For during our observations, comparisons between the observed circularly polarized flux with the theoretical flux from pure cyclotron models was required. Since we detected negative polarization, we took the absolute value of the polarized flux. 
We then modelled the total circularly polarized flux following the cyclotron emission models from the stratified accretion shocks as described in \cite{1998PhDT.......159P}. 
The results, for phase = 0.91, before the eclipse, are shown in Fig. \ref{fig:circ_model} and we have over-plotted a pure cyclotron model with the magnetic field of 57 MG viewed at an angle of 70$^{\circ}$ to the line of sight, consistent with that given in \citet{1990A&A...230..120S}. 
The three cyclotron features mentioned above may be identified with harmonic numbers 4, 3 and 2. 
We also used Equ. \ref{equ:harm} to determine the strength of the magnetic field and utilizing the angle $\theta$ = 70$^{\circ}$ mentioned above, this corresponds to the mean magnetic field of $\sim$58 MG. As is evident from the figure, not all the harmonics can be described by the model. 
Also, it is not possible to fit all the harmonics seen with the single value of the magnetic field.  

\subsection{MeerKAT radio results}\label{sect:radio_res}

\begin{figure*}
    \centering
    \includegraphics[width = \textwidth]{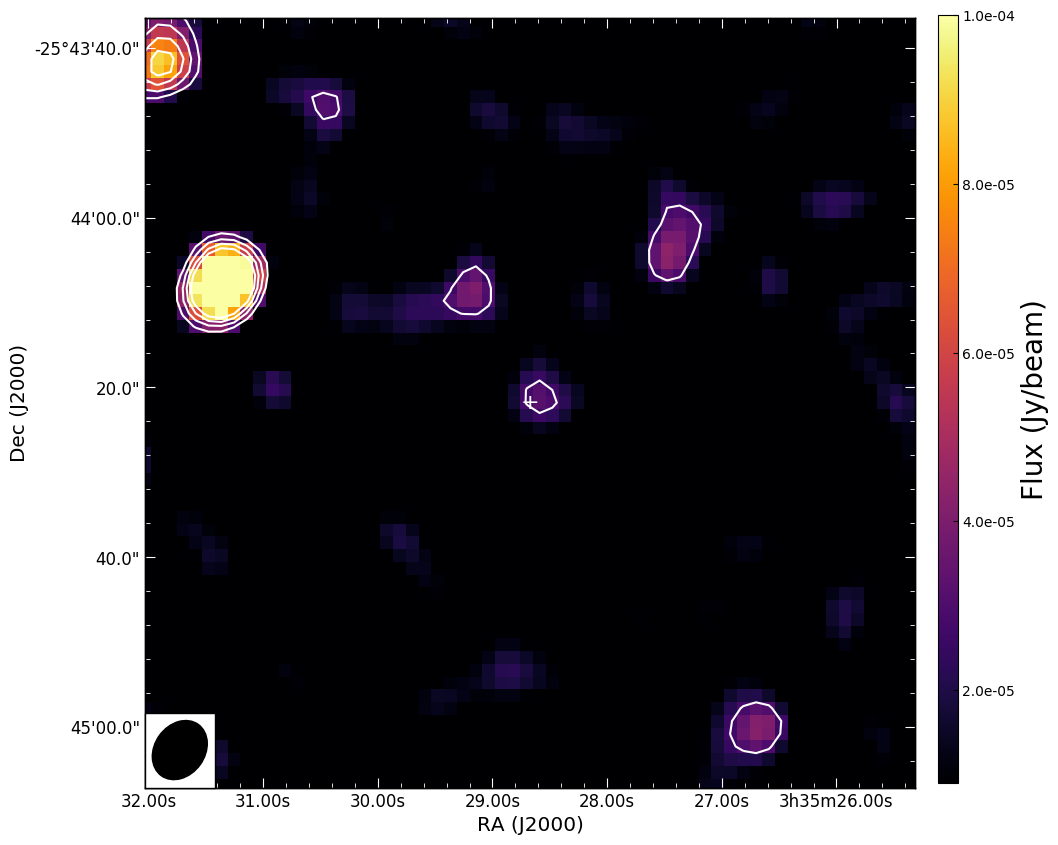}
    \caption{Colour map with contours overlayed of the field surrounding UZ For. The contours are at 3$\sigma$, 5$\sigma$, 7$\sigma$ and 9$\sigma$ levels. The beam on the lower left corner has dimensions of 7.45$''$ $ \times$ 5.91$''$ and a position angle of -35.81$^{\circ}$. The plus (+) sign at the centre of the image marks the position of the optical coordinates of UZ For.}
    \label{fig:radio}
\end{figure*}

Figure \ref{fig:radio} shows the field surrounding UZ For at radio. The position of the optical coordinates for UZ For is indicated with a plus (+) sign at the centre of the image. There are some noticeable radio sources in the field of view. 
Our observations taken with the MeerKAT telescope in imaging mode show a faint source, with a peak flux of 30.7$\pm$5.4 $\mu$Jy/beam (3.4$\sigma$), located at (epoch J2000) RA: 03:35:28.596 $\pm$ 0.024 and Dec: -25:44:21.331 $\pm$ 0.344. The rms noise is estimated to be 9 $\mu$Jy/beam. The synthesized beam size is 7.45 $\times$ 5.91 arsec$^2$ at a position angle of -35.81$^{\circ}$. 
The position of this source coincides, within the uncertainty given, with the optical coordinates of UZ For -- RA: 03:35:28.652 $\pm$ 0.048 and Dec: -25:44:21.766 $\pm$ 0.057 (epoch J2000, \citealt{2018yCat.1345....0G}).


\section{Discussion and conclusions}\label{sec:dis}

We have presented the phase-resolved spectroscopy and circular spectrospolarimetry obtained with the SALT telescope as well as photopolarimetry and radio observations of UZ For.

\subsection{Spectroscopy and Doppler tomography}

The blue averaged spectrum of UZ For is dominated by strong emission from the Balmer lines and HeII 4686\AA{} with weak emission from the HeI lines and the Bowen blend. 
The strength of the Balmer lines, HeI lines and HeII 4686\AA{} are consistent with the low resolution spectra presented by \citet{1989ApJ...337..832F} obtained when UZ For was in high state. This suggest that UZ For was observed in the high-state in 2013 January. The ratio between H$\beta$ to H$\gamma$ or H$\delta$ is close to unity, signifying that these lines are emitted in an optically thick region. 
The spectra of polars in high states consists entirely of emissions lines superimposed on a steep continuum that rises strongly towards the red or the blue. For example, BL Hyi was observed in high-state by \cite{2006PASP..118..678G} and its spectrum show strong emission lines with a continuum that rises in the blue. 
The red averaged spectrum shows weak emission from the irradiated face of the secondary star, e.g. CaII lines at 8498 and 8542\AA{}. 
Our averaged spectrum of UZ For shows a flat Balmer decrement. This is not consistent with the steep Balmer decrement\footnote{The spectrum in question here was averaged from phase 0.2 to 0.6 -- to exclude contribution from the magnetic pole.} reported by \cite{1989ApJ...347..426A}. 

We presented the first detailed Doppler tomographic analysis of UZ For. Our observed trailed spectra in the blue show three distinct emission components; 1) a relatively narrow component with low-velocity amplitude, 2) a broad emission line which has a high velocity amplitude, and 3) a relatively broad feature which is visible throughout the orbital phase. 
In the red, the observed trailed spectra of CaII 8542 \AA{} shows a major component which is associated with the irradiated secondary star. The basic structure of the observed trailed spectra is reproduced in the reconstructed trailed spectra.

The various emission components seen in the trailed spectra are reproduced in the Doppler maps (Figs \ref{figure:doppsBeta_A} to \ref{figure:doppsCaII8542a}, top rows) and more specifically in the inside-out tomogram. In the standard projection it is difficult to distinguish the various emission components seen in the trailed spectra since the ballistic stream and the threading region dominates the emission. 
Our Doppler map results of the H$\beta$  and HeII 4686 \AA{} lines are consistent with those presented by \citet{1999ASPC..157...71S} for HeII 4686\AA{}. 

Our modulation amplitude maps (Figs \ref{figure:doppsBeta_B} and \ref{figure:doppsHeII_B}, top panels) show that at least two emission components are flux modulated: the ballistic and the magnetic confined accretion streams, are obviously modulated in both tomograms. The Doppler map based in the inside-out projection shows that the secondary star is also flux modulated. This is not clear in the Doppler map based on the standard projection.    

The Doppler maps presented here are dominated by emission from the ballistic and  magnetic confined accretion stream. This is not the case for HU Aqr, which shows Doppler maps that are dominated by the emission from the irradiated face of the secondary star and the ballistic stream \citep{2016A&A...595A..47K}. 
The same author showed the Doppler maps of V834 Cen were dominated by emission from the ballistic and magnetic accretion stream, like for this UZ For study.   

\subsection{Photopolarimetry}

In Sect. \ref{sect:polar} we presented both circular and linear photopolarimetric observations of UZ For. The circular polarization results show that UZ For is negatively polarized from phases $\sim$0.7--1.1 with polarization reaching $\sim$-5\% before the eclipse. 
Our results are not consistent with those reported in literature by \citet{1988ApJ...329L..97B} and \citet{1989ApJ...337..832F} in that we see an increase in negative circular polarization from phases greater than 0.6 leading to the eclipse. These authors reported positive polarization at phases beginning at 0.65 and lasting until phase 1.15. 
The reversal in the sign of polarization seen in our results suggest that during our observations the second pole (in the opposite hemisphere) is responsible for the polarized radiation. VV Pup is an example of a polar that have been found to show negative polarization \citep{1979ApJ...229..652L}. This was interpreted as evidence that the additional radiation is due to accretion onto a second region of the WD, where the longitudinal component of the magnetic field is of the opposite sign with respect to that at the primary accreting pole. 

We also presented linear polarization of UZ For and the results shows polarization reaching $\sim$5\% leading to the eclipse. A pulse of linear polarization is seen just after the eclipse reaching $\sim$10\% and decreasing gradually before phase 1.1. 
After that, the polarization is consistent with 5\%. The percentage of linear polarization is consistent with the results presented by \cite{1988ApJ...329L..97B}. 

\subsection{Spectropolarimetry}

Our spectropolarimetry results show a continuum that rises in the blue with a broad hump around 5500 \AA{}. The slope of the continuum is consistent with those recorded in the literature when UZ For was in a low state \citep{1988A&A...195L..15B,1990A&A...230..120S}. 
The only difference between our spectra and theirs is the strength of the emission lines. 
Our spectra was taken during a high state and this is supported by strong emission from the HeII and Balmer lines.  

Our polarized spectra show negative circular polarization reaching $\sim$-8\% in the blue before the eclipse and decreasing with increasing wavelength. The circularly polarized spectra showed three cyclotron harmonics associated with harmonic numbers 4, 3 and 2, respectively. These three features weaken going into the eclipse. 
When the WD and the accretion spot(s) are emerging from the eclipse, only the strongest negative polarized hump ($\sim$4500 \AA{}) is still visible and there is a possibility of an additional hump below 4000 \AA{} but this appears positive. The additional hump is present in the spectra presented by \cite{1989ApJ...337..832F}. After the eclipse, the spectrum appears flat and exhibit nonzero polarization in the red.  

The resulting polarized flux from this cyclotron spectra was modelled using pure cyclotron models with the magnetic field strength of 57 MG. 
We note that our overlayed model does not fit all the humps well especially in the red part of the spectrum. We attribute this to the second spot or pole on the surface of the WD also emitting cyclotron radiation. The model used is specific for a given magnetic field and depends on other parameters like the electron temperatures, optical depth, viewing angle, etc., so changing any of these parameters will give us a slightly different fit to the flux. 
Previous studies of UZ For in low state  \citep{1988A&A...195L..15B,1989ApJ...347..426A,1990A&A...230..120S} and high state \citep{1989ApJ...337..832F} revealed the presence of cyclotron humps at different wavelength and these were attributed to the field strength of 53--57 MG for the main accretion region with the possibility of the second pole of 33--75 MG emitting cyclotron radiation as well. 
The position of the harmonics presented in our work is not consistent with those of the other authors. This is expected since position of the harmonics depends on the electron temperature, optical depth and viewing angle. The strength of the magnetic field derived in our work is consistent with those of \citet{1990A&A...230..120S}.

\subsection{Radio Emission}

We detected radio emission at the expected position of UZ For using MeerKAT in the L-band centred at 1.28 GHz with a peak flux  of 30.7$\pm$5.4 $\mu$Jy/beam. 
The reported magnitude of UZ For around the time of MeerKAT observations by the AAVSO ranges between 16.5-16.1. This is consistent with the out-of-eclipse $i,q,r,g,z$ magnitudes obtained from the MeerLICHT observations. 
Recently, \citet{2017AJ....154..252B} detected UZ For in the radio using the VLA at C-band (4-6 GHz) and they found it to have a flux density of 315 $\pm$ 101 $\mu$Jy. 
The other bands, X-band (8-10 GHz) and K-band (18-22 GHz), yielded nondetections of the source. 
Our flux density for this source at 1.28 GHz is ten times fainter than previously recorded at 4-6 GHz, and demonstrates the sensitivity of the MeerKAT telescope. 
Our results suggest that UZ For is variable in radio wavelengths but the time-scales of this source's variability is not yet known. The majority of the mCVs studied by \citet{2017AJ....154..252B} showed radio emission in no more than two frequency band (except AM Her) and epochs (except AR UMa and AE Aqr). 
UZ For lies in one of the MIGHTEE fields \citep{2016mks..confE...6J} and therefore we will continue monitoring it in optical and radio wavelengths.

\section*{Acknowledgements}

The spectroscopic observations reported in this paper were obtained with the Southern African Large Telescope (SALT) in the facilities of the SAAO in Sutherland under programs 2012-2-RSA-008 and 2013-2-RSA-006 (PI: Stephen. B. Potter) and 2018-2-LSP-001 (PI: David Buckley). 
We thank the staff at the South African Radio Astronomy Observatory (SARAO) for scheduling these observations. 
The MeerKAT telescope is operated by the South African Radio Astronomy Observatory, which is a facility of the National Research Foundation (NRF), an agency of the Department of Science and Innovation. 
This work was carried out in part using facilities and data processing pipelines developed at the Inter-University Institute for Data Intensive Astronomy (IDIA). IDIA is a partnership of the Universities of Cape Town, of the Western Cape and of Pretoria. 

The financial assistance of the NRF towards this research is hereby acknowledged. 
ZNK acknowledges funding by the NRF and the University of Cape Town (UCT) through a PhD bursary. 
PAW acknowledges the NRF and the UCT for their financial support. Part of this work was supported under the BRICS STI framework programme (South African grant UID110480, Russian grant RFFI 17-52-80139). 
KP acknowledges funding by the National Astrophysics and Space Science Programme (NASSP), the NRF of South Africa through a SARAO bursary, and the UCT for work on MeerLICHT. 


\bibliographystyle{mnras}
\bibliography{References_uzfor}
\bsp	
\label{lastpage}
\end{document}